\documentclass[iop]{emulateapj}
\usepackage{graphicx}
\usepackage[normalem]{ulem}

\def\gsim{\mathrel{\raise0.35ex\hbox{$\scriptstyle >$}\kern-0.6em\lower0.40ex\hbox{{$\scriptstyle \sim$}}}}
\def\lsim{\mathrel{\raise0.35ex\hbox{$\scriptstyle <$}\kern-0.6em\lower0.40ex\hbox{{$\scriptstyle \sim$}}}}

\begin{document}

\title{ALMA reveals potential evidence for spiral arms, bars, and rings in high-redshift submillimeter galaxies} 

\shorttitle{Sub-kpc dust structure in high-redshift SMGs}
\shortauthors{Hodge et al.}

\author{J. A. Hodge\altaffilmark{1}}
\altaffiltext{1}{Leiden Observatory, Leiden University, P.O. Box 9513, 2300 RA Leiden, the Netherlands}
\email{hodge@strw.leidenuniv.nl}

\author{I. Smail\altaffilmark{2,3}}
\altaffiltext{2}{Centre for Extragalactic Astronomy, Department of Physics, Durham University, South Road, Durham, DH1 3LE, UK}
\altaffiltext{3}{Institute for Computational Cosmology, Durham University, South Road, Durham, DH1 3LE, UK}

\author{F. Walter\altaffilmark{4}}
\altaffiltext{4}{Max--Planck Institut f\"ur Astronomie, K\"onigstuhl 17, 69117 Heidelberg, Germany}

\author{E. da Cunha\altaffilmark{5}}
\altaffiltext{5}{Research School of Astronomy and Astrophysics, Australian National University, Canberra, ACT 2611, Australia}

\author{A. M. Swinbank\altaffilmark{2,3}}

\author{M. Rybak\altaffilmark{1}}

\author{B. Venemans\altaffilmark{4}}

\author{W. N. Brandt\altaffilmark{6,7,8}}
\altaffiltext{6}{Department of Astronomy \& Astrophysics, 525 Davey Lab, The Pennsylvania State University, University Park, Pennsylvania, 16802, USA}
\altaffiltext{7}{Institute for Gravitation and the Cosmos, The Pennsylvania State University, University Park, PA 16802, USA}
\altaffiltext{8}{Department of Physics, 104 Davey Laboratory, The Pennsylvania State University, University Park, PA 16802, USA}

\author{G. Calistro Rivera\altaffilmark{1}}

\author{S. C. Chapman\altaffilmark{9}}
\altaffiltext{9}{Department of Physics and Atmospheric Science, Dalhousie University, 6310 Coburg Road, Halifax, NS B3H 4R2, Canada}

\author{Chian-Chou Chen\altaffilmark{10}}
\altaffiltext{10}{European Southern Observatory, Karl Schwarzschild Strasse 2, Garching, Germany}

\author{P. Cox\altaffilmark{11}}
\altaffiltext{11}{Institut d'Astrophysique de Paris, CNRS, UMR 7095, 98bis boulevard Arago, 75014, Paris, France}

\author{H. Dannerbauer\altaffilmark{12,13}}
\altaffiltext{12}{Instituto de Astrof\'{\i}sica de Canarias, V\'{\i}a L\'actea s/n, 38205, La Laguna, Tenerife, Spain}
\altaffiltext{13}{Universidad de La Laguna, Dpto. Astrof'sica, E-38206 La Laguna, Tenerife, Spain}

\author{R. Decarli\altaffilmark{14}}
\altaffiltext{14}{INAF / Osservatorio di Astrofisica e Scienza dello Spazio di Bologna, Via Gobetti 93/3, 40129 Bologna, Italy}

\author{T. R. Greve\altaffilmark{15,16}}
\altaffiltext{15}{Cosmic Dawn Center (DAWN), DTU-Space, Technical University of Denmark, Elektrovej 327, DK-2800 Kgs. Lyngby; Niels Bohr Institute, University of Copenhagen, Juliane Maries Vej 30, DK-2100 Copenhagen $\O$} 
\altaffiltext{16}{University College London, Department of Physics \& Astronomy, Gower Street, London, WC1E 6BT, UK}

\author{K. K. Knudsen\altaffilmark{17}}
\altaffiltext{17}{Department of Space, Earth and Environment, Chalmers University of Technology, Onsala Space Observatory, SE--43992 Onsala, Sweden}

\author{K. M. Menten\altaffilmark{18}}
\altaffiltext{18}{Max--Planck Institut f\"ur Radioastronomie, Auf dem H\"ugel 69, D--53121 Bonn, Germany}

\author{E. Schinnerer\altaffilmark{5}}

\author{J. M. Simpson\altaffilmark{19}}
\altaffiltext{19}{Academia Sinica Institute of Astronomy and Astrophysics, No. 1, Sec. 4, Roosevelt Rd., Taipei 10617, Taiwan}

\author{P. van der Werf\altaffilmark{1}}

\author{J. L. Wardlow\altaffilmark{20}}
\altaffiltext{20}{Physics Department, Lancaster University, Lancaster, LA1 4YB, UK}

\author{A. Weiss\altaffilmark{18}}

%%%%%%%%%%%%%%%%%%%%%%%%%%%%%%%%%%%%%%%%%%%%%%%
\begin{abstract}
\noindent We present sub-kpc-scale mapping of the 870\,$\mu$m ALMA continuum emission in six luminous ($L_{\rm IR}~\sim~5~\times10^{12}$\,L$_{\odot}$) submillimeter galaxies (SMGs) from the ALESS survey of the Extended \textit{Chandra} Deep Field South. Our high--fidelity 0.07$''$-resolution imaging ($\sim$500\,pc) reveals robust evidence for structures with deconvolved sizes of $\lesssim$0.5--1\,kpc embedded within (dominant) exponential dust disks. The large-scale morphologies of the structures within some of the galaxies show clear curvature and/or clump-like structures bracketing elongated nuclear emission, suggestive of bars, star-forming rings, and spiral arms. 
In this interpretation, the ratio of the `ring' and `bar' radii (1.9$\pm$0.3) agrees with that measured for such features in local galaxies. These potential spiral/ring/bar structures would be consistent with the idea of tidal disturbances, with their detailed properties implying flat inner rotation curves and Toomre-unstable disks ($Q<1$). The inferred one-dimensional velocity dispersions ($\sigma_{\rm r}\lesssim$ 70--160\,km s$^{-1}$) are marginally consistent with the limits implied if the sizes of the largest structures are comparable to the Jeans length. 
We create maps of the star formation rate density ($\Sigma_{\rm SFR}$) on $\sim$500\,pc scales and show that the SMGs are able to sustain a given (galaxy-averaged) $\Sigma_{\rm SFR}$ over much larger physical scales than local (ultra--)luminous infrared galaxies. However, on 500\,pc scales, they do not exceed the Eddington limit set by radiation pressure on dust. If confirmed by kinematics, the potential presence of non-axisymmetric structures would provide a means for net angular momentum loss and efficient star formation, helping to explain the very high star formation rates measured in SMGs.


\end{abstract}

%%%%%%%%%%%%%%%%%%%%%%%%%%%%%%%%%%%%%%%%%%%%%%%
\section{INTRODUCTION}
\label{Intro}

At the peak of the cosmic star formation rate density ($z\sim2$), the majority of the star formation in the Universe occurred behind dust \citep[e.g.,][]{Madau2014}. This has made it difficult to obtain a complete picture of galaxy evolution, particularly for the most actively star-forming population, which can be rendered faint or even invisible in the dust-sensitive rest-frame optical/UV imaging \citep[e.g.,][]{Walter2012}. In these galaxies, the majority of the rest-frame optical/UV light is re-radiated in the far-infrared (FIR), resulting in large submillimeter flux densities for the high-redshift sources. 
Although such `submillimeter-selected galaxies' \citep[SMGs; e.g.,][]{2002PhR...369..111B, Casey2014} have been known about for over twenty years -- and although they have been shown to contribute significantly to the cosmic star formation rate density \citep[e.g.,][]{Swinbank2014} -- there is still considerable uncertainty over their detailed physical properties and overall nature.

\begin{deluxetable*}{ l c c c c c }
\tabletypesize{\small}
%\tablewidth{16cm}
\tablecaption{Galaxy properties}
\tablehead{
\colhead{Source ID$^{a}$} & \colhead{$z^b$} & \colhead{$z_{\rm source}$$^{b}$} & \colhead{log(M$_*$/M$_{\odot}$)$^{c}$} & \colhead{log(SFR/M$_{\odot}$ yr$^{-1}$)$^{c}$} & \colhead{T$_{\rm dust}$/K$^{c}$}  }
 \startdata
 ALESS 3.1 & 3.374 & CO(4--3) & 11.30$^{+0.19}_{-0.24}$ & 2.81$^{+0.07}_{-0.08}$ & 36$^{+5}_{-2}$\\ 
 ALESS 9.1 & 4.867 & CO(5--4) & 11.89$^{+0.12}_{-0.12}$ & 3.16$^{+0.07}_{-0.08}$ & 51$^{+5}_{-4}$ \\ 
 ALESS 15.1 & 2.67 & $z_{\rm phot}$ & 11.76$^{+0.21}_{-0.26}$ & 2.44$^{+0.15}_{-0.26}$ & 33$^{+7}_{-4}$\\ 
 ALESS 17.1 & 1.539 & H$\alpha$, CO(2--1) & 11.01$^{+0.08}_{-0.07}$ & 2.29$^{+0.02}_{-0.03}$ & 28$^{+6}_{-0}$\\
 ALESS 76.1 & 3.389 & [OIII] & 11.08$^{+0.29}_{-0.34}$ & 2.56$^{+0.11}_{-0.12}$ & 37$^{+10}_{-4}$\\
 ALESS 112.1 & 2.315 & Ly$\alpha$ & 11.36$^{+0.09}_{-0.12}$ & 2.40$^{+0.07}_{-0.08}$ & 31$^{+5}_{-2}$
 \enddata
 \tablecomments{$^{a}$ Source IDs are from \citet{Hodge2013a}.\\
 $^{b}$ Rest-frame optical/UV-based spectroscopic redshifts are from \citet{Danielson2017}, CO-based redshifts are from Weiss et al.\,(in prep) or Wardlow et al.\,(in prep), and the photometric redshift was taken from \citet{daCunha2015}.\\
 $^{c}$ Stellar masses, SFRs, and luminosity-averaged dust temperatures are from multi-wavelength SED fits which were updated from those presented in \citet{daCunha2015} to include new ALMA band 4 data (da Cunha et al.\,in prep.). In cases where an updated redshift was available, they were recalculated using the same method.\\ 
}
 \label{tab:physprops}
 \end{deluxetable*}

The recent advent of the Atacama Large Millimeter Array (ALMA) is providing unique insights into high-redshift dusty star formation. In particular, the combination of ALMA's unprecedented sensitivity and resolution has allowed for spatially resolved (i.e., sub-galactic) studies of the rest-frame FIR emission in the SMG population \citep[e.g.,][]{Simpson2015b, Simpson2017, Ikarashi2015, Hodge2016, Chen2017, CalistroRivera2018, Fujimoto2018}, sometimes at even higher-resolution than is possible in the optical \citep[$\sim$0.03$''$; e.g.,][]{Iono2016, Oteo2017, Gullberg2018}. While there is still debate over where SMGs lie relative to the SFR--mass trend \citep[e.g.,][]{daCunha2015, Koprowski2016, Danielson2017, Elbaz2018}, one thing that is becoming clear in all of these studies is that the distribution of dusty star formation (traced by the rest-frame FIR emission) is relatively compact ($\sim$3$\times$ smaller) compared to the rest-frame optical/UV emission visible with the \textit{Hubble Space Telescope} \citep[\textit{HST}; e.g.,][]{Chen2015, Simpson2015b, CalistroRivera2018}, and that it is disk-like on galaxy-wide scales \citep[S\'ersic index $n\sim1$; e.g.,][]{Hodge2016}.   

There have been varying reports on whether the rest-frame FIR emission traced by ALMA submillimeter continuum observations shows evidence for structure on sub-galactic scales. While some studies report evidence that a fraction of the submillimeter emission from some SMGs breaks up into `clumps' on sub-kpc or even kpc scales \citep[e.g.,][]{Iono2016, Oteo2017}, other studies find that the bulk of the observed emission is consistent with smooth disk emission given the signal-to-noise \citep[e.g.,][]{Hodge2016, Gullberg2018}. Clumpy emission has been claimed previously on these scales based on observations of kpc-scale UV clumps in high-redshift galaxies \citep[e.g.,][]{Dekel2009b, ForsterSchreiber2011, Guo2012, Guo2015}, although there is little evidence these represent true structures in the molecular gas or dust in these galaxies.

If the intense starbursts ($\sim$100 to $>$1000 M$_{\odot}$ yr$^{-1}$) observed in SMGs are triggered by galaxy interactions/mergers, as is commonly believed, then we might also expect to see morphological evidence of these interactions/mergers. In particular, it has long been known from early numerical work  \citep[e.g.,][]{Noguchi1987} that tidal disturbances can induce the formation of non-axisymmetric features such as galactic bars and spiral arms. Simulations suggest that spirals of the $m=2$ variety (i.e., double-armed) are actually difficult to produce \textit{except} through tidal interactions/bars \citep{Kormendy1979, Bottema2003}, with the most prominent grand-design spiral arms appearing in interacting galaxies such as M51. While the efficiency of their formation depends on the exact details of the orbital path and mass ratio \citep[e.g.,][]{Athanassoula2003, Lang2014, Kyziropoulos2016, Gajda2017, Pettitt2018}, these non-axisymmetric features can have significant consequences for the galactic dynamics. Specifically, they can interact with galactic material and cause resonances, including the corotation and inner and outer Lindblad resonances \citep{Sellwood1993}. Gas accumulates at these resonances and produces star-forming rings \citep[e.g.,][]{Schwarz1981, Buta1986, Buta1996, Rautiainen2000}. More critically, non-axisymmetric features such as bars can also efficiently redistribute the angular momentum of the baryonic and dark matter components of disk galaxies \citep[e.g.,][]{Weinberg1985, Athanassoula2002, Marinova2007}, triggering gas inflow and nuclear starbursts and thus driving spheroid growth.
	
The physical processes that accompany the intense bursts of star formation seen in systems such as SMGs and ultra-luminous infrared galaxies (ULIRGs) are also thought to create feedback on the star-forming gas, potentially even slowing or halting further gravitational collapse in a self-regulating process. In particular, radiation pressure from massive stars on dust (which is coupled to the gas through collisions and magnetic fields) may play an important role in regulating star formation in the optically thick centers of starbursts like local ULIRGs \citep{Scoville2003, Murray2005, Thompson2005, Andrews2011}, where almost all of the momentum from the starlight is efficiently transferred to the gas. Indeed, \citet{Thompson2005} showed that radiation pressure could make up the majority of the vertical pressure support in so-called `Eddington-limited' dense starbursts.

While the latest ALMA results show that most SMGs are not approaching the Eddington limit for star formation on galaxy-wide scales \citep[e.g.,][]{Simpson2015b}, this does not mean that the star formation is not limited by radiation pressure on more local (kpc or sub-kpc) scales, as has been observed in more compact local ULIRGs \citep[e.g.,][]{BarcosMunoz2017} or even for giant molecular clouds in our own Milky Way \citep[e.g.,][]{Murray2010, Murray2011}. 
Similarly, while the bulk of the submillimeter emission in SMGs appears to be arising from a disk-like distribution on $\gtrsim$kpc scales, this does not mean that these dust and gas disks are featureless. In answering these open issues, obtaining higher angular resolution does not necessarily help unless one has correspondingly better surface brightness sensitivity to map the significance of beam-sized features with adequate S/N \citep[e.g.,][]{Hodge2016}.

In this work, we present high-resolution ($\sim$0.07$''$), high-fidelity ALMA imaging of the submillimeter emission (rest-frame FIR emission) from six SMGs at redshifts $1.5<z<4.9$ from the ALMA follow-up of the LABOCA ECDFS submillimeter survey \citep[ALESS;][]{Hodge2013a}, allowing us to study the morphology and intensity of their dusty star formation on $\sim$500\,pc scales. We present the details of the observations and data reduction in \S\ref{data}.  The results are presented in \S\ref{results}, including a comparison with \textit{HST} imaging (\S\ref{HSTcomp}), an analysis of the sub-kpc structure (\S\ref{structure}), the presentation of SFR density maps (\S\ref{SFRDmaps}), and a comparison to the SFR--mass trend (\S\ref{MScomp}). \S\ref{discussion} presents a discussion of these results, followed by a summary of the conclusions in \S\ref{summary}. Throughout this work, we assume a standard $\Lambda$CDM cosmology with H$_0$=67.8 km\,s$^{-1}$ Mpc$^{-1}$, $\Omega_{\Lambda}$=0.692, and $\Omega_{M}$=0.308 \citep{Planck2016}.

%%%%%%%%%%%%%%%%%%%%%%%%%%%%%%%%%%%%%%%%%%%%%%%%
\section{Observations and Data Reduction}
\label{data}

\subsection{ALMA Sample Selection \& Observations}
\label{obs}

The ALMA observations presented here were taken in six observing blocks from 28 July to 27 Aug 2017 as part of project \#2016.1.00048.S. In order to maximize S/N for the high-resolution observations requested, the six SMGs were selected as the submillimeter-brightest sources from the 16 ALESS SMGs with previous high-resolution (0.16$''$) 870\,$\mu$m ALMA imaging from \citet{Hodge2016}, which were themselves chosen as the submillimeter-brightest sources with (randomly-targeted) \textit{HST} coverage. All of the sources have existing \textit{HST} data from CANDELS or our own program \citep{Chen2015}. No pre-selection was made on morphology/scale of the emission in the previous ALMA or \textit{HST} imaging so as to avoid biasing the results.  

The observations were carried out in an extended configuration, with a maximum baseline of 3.7\,km. The average number of antennas present during the observations was 45 (with a range of 42--47). The 5th percentile of the baseline uv-distances of the delivered data is 200\,m, giving a maximum recoverable scale (MRS) of 0.9$''$ according to Equation 7.7 of the ALMA Cycle~4 Technical Handbook. This corresponds to a physical scale of $\sim$7.5\,kpc at a redshift of $z\sim2.5$.

With the aim of quantifying the emission potentially resolved out by the requested extended-configuration observations, we utilized a spectral setup identical to the original Cycle~0 ALESS observations of these galaxies \citep{Hodge2013a} as well as the subsequent 0.16$''$ observations by \citet{Hodge2016}. This setup centered at 344 GHz (870\,$\mu$m) with 4$\times$128 dual polarization channels covering the 8 GHz bandwidth. We utilized ALMA's Band 7 in Time Division Mode (TDM). At the central frequency, the primary beam is 17.3$''$ (FWHM). The total on-source time for each of the science targets was approximately 50 minutes, and we requested standard calibration. The median precipitable water vapor at zenith ranged from 0.4--1.0\,mm across the six datasets, with an average value of 0.5\,mm.

Due to the selection criteria, the targets of this paper are some of the submillimeter-brightest sources of the ALESS SMG sample as a whole \citep[Table~\ref{tab:fluxcomp};][]{Hodge2013a}. They have redshifts that range from $\sim$1.5--4.9 (Table~\ref{tab:physprops}), including five derived from optical and submillimeter spectroscopy \citep[Weiss et al.\,in prep.]{Danielson2017} and one from photometry \citep{daCunha2015}. Their median redshift ($z=3.0\pm0.5$) is consistent with the full ALESS sample \citep[$z=2.7\pm0.1$;][]{daCunha2015}. 
Their stellar masses, star formation rates, and dust temperatures were derived from multiwavelength SED fits, which were updated from those presented in \citet{daCunha2015} to include new ALMA Band 4 data (da Cunha et al.\,in prep.).
Their median star formation rate ($\sim$300 M$_{\odot}$ yr$^{-1}$) is consistent with the ALESS sample as a whole \citep{Swinbank2014, daCunha2015}, while their median dust temperature (34$\pm$3 K) is marginally cooler than the full sample as analyzed by \citet{daCunha2015}. Their median stellar mass ($\sim$2$\times$10$^{11}$ M$_{\odot}$) is also larger than the median of the full sample \citep[$\sim$8$\times$10$^{10}$ M$_{\odot}$;][]{Simpson2014}, indicating that we may be probing the high--mass end of the population. One of the six sources is associated with an X-ray source and is classified as an AGN \citep[ALESS 17.1, $L_{\rm 0.5-8keV,corr}$ $=$ 1.2 $\times$ 10$^{43}$\,ergs s$^{-1}$;][]{Wang2013}.

\begin{figure*}
\centering
\includegraphics[scale=0.8,trim={0 2cm 0 1cm},clip]{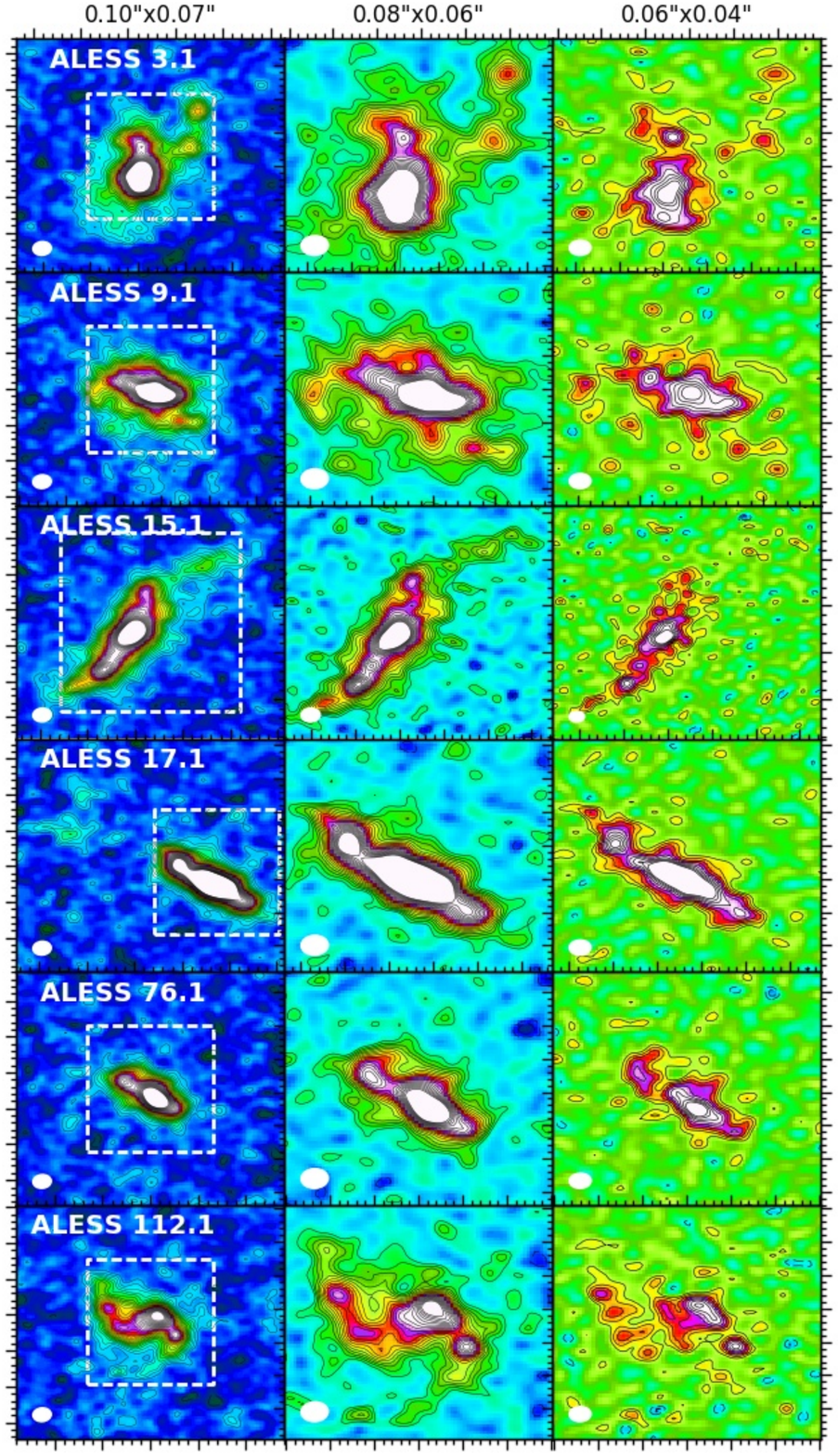}
\caption{ALMA maps of the 870\,$\mu$m continuum emission from six SMGs imaged at three different resolutions (indicated above each column). Contours start at $\pm$2$\sigma$ and go in steps of 1$\sigma$, stopping at 30$\sigma$ (left), 20$\sigma$ (middle), and 10$\sigma$ (right) for clarity. These images reveal resolved structure on scales of $\sim$0.07$''$ ($\sim$500\,pc at $z\sim2.5$), with large-scale structures suggestive of spiral arms and bars. 
\textit{Left column:} 1.3$''$$\times$1.3$''$ maps imaged with natural weighting, resulting in an RMS of $\sigma$$\sim$20\,$\mu$Jy beam$^{-1}$ and a resolution of 0.10$''$$\times$0.07$''$. The dashed white box indicates the region shown in the two right columns and is 0.7$''$$\times$0.7$''$ for all sources except ALESS 15.1, where a larger 1.0$''$$\times$1.0$''$ region is shown.
\textit{Middle column:} Zoomed-in maps of the region indicated in the left column, now imaged with Briggs weighting ($R = +0.5$), resulting in an RMS of $\sigma$$\sim$22\,$\mu$Jy beam$^{-1}$ and a resolution of 0.08$''$$\times$0.06$''$. 
\textit{Right column:} Zoomed-in maps of the region indicated in the left column, now imaged with a different Briggs weighting ($R = -0.5$), resulting in an RMS of $\sigma$$\sim$42\,$\mu$Jy beam$^{-1}$ and a resolution of 0.06$''$$\times$0.04$''$.  }
\label{fig:ALESScyc4_ALL}
\end{figure*}

%%%%%%%%%%%%%%%%%%%%%%%%%%%%%%%%%%%%%%%%%%%%%%%
\subsection{ALMA Data Reduction \& Imaging }
\label{reduction}

The ALMA data were reduced and imaged using the Common Astronomy Software Application\footnote{http://casa.nrao.edu} ({\sc casa}) version 4.7. 
Inspection of the pipeline-calibrated data tables revealed data of high quality, and the $uv$-data were therefore used without further modification to the calibration scheme or flagging.

Prior to imaging, the data were combined with the lower-resolution ($\sim$0.16$''$), lower-sensitivity data previously obtained for these sources at the same frequency and presented in \citet{Hodge2016}. Due to the lower sensitivity of the previous data, as well as the large maximum recoverable scale (MRS) already achieved by the new data (\S\ref{obs}), this made very little difference to the resulting image quality.

Imaging of the combined data was done using {\sc casa}'s {\sc clean} task and multi-scale {\sc clean}, a scale-sensitive deconvolution algorithm \citep{Cornwell2008}. For this we employed a geometric progression of scales, as recommended, and we found that the exact scales used did not affect the outcome. The use of multi-scale {\sc clean} made little qualitative difference to the final images, in comparison to those imaged without multi-scale {\sc clean}, but we found that the residual image products from the runs without multi-scale {\sc clean} showed a significant plateau of positive uncleaned emission that was absent in the residual maps made with multi-scale {\sc clean}. 
We therefore use the multi-scale {\sc clean} results for the remainder of the analysis.

Cleaning was done interactively by defining tight clean boxes around the sources and cleaning down to 1.5$\sigma$. Different weighting schemes were utilized on the \textit{uv}-data in order to produce images at different spatial resolutions and thus investigate the structure in the sources. As a point of reference, imaging the data with Briggs weighting \citep{Briggs1999} and a robust parameter of $R = +0.5$ -- generally a good compromise between resolution and sensitivity -- produced images with a synthesized beam size of 0.08$''$$\times$0.06$''$ and a typical RMS noise of 23\,$\mu$Jy beam$^{-1}$. With this array configuration and source SNR, the astrometric accuracy of the ALMA data is likely limited by the phase variations over the array to a few mas.\footnote{ALMA Cycle~5 Technical Handbook, Chapter 10.6.6.}

The MRS of the newly delivered data (0.9$''$; \S\ref{obs}) is larger than the median major axis FWHM size of the ALESS sources at this frequency \citep[0.42$''$$\pm$0.04$''$;][]{Hodge2016}, indicating that most of the flux density should be recovered. To test this, we $uv$-tapered the concatenated data to 0.3$''$, cleaned them interactively, and measured the integrated flux densities, as the sources are still resolved at this resolution. 
The results are shown in Table~\ref{tab:fluxcomp} along with the flux densities measured from the compact-configuration ($\sim$1.6$''$) Cycle~0 observations \citep{Hodge2013a}. In general, we recover most of the flux density measured in the lower-resolution Cycle~0 observations, indicating that the sources are relatively compact. For two of the six sources, the current data may be missing $\sim$20\% of the total 870\,$\mu$m emission, indicating the presence of a low-surface-brightness and/or extended component to the emission not recoverable in the present data. We therefore report any fractional contributions from structures detected in this work using the total flux densities derived in the lower-resolution Cycle~0 observations.

\begin{deluxetable*}{ l c c c }[]
\tabletypesize{\small}
%\tablewidth{10cm}
\tablecaption{870\,$\mu$m continuum properties}
\tablehead{
\colhead{Source ID} & \colhead{Cycle~0 (1.5$''$)} & \colhead{This work (0.3$''$ taper)} & \colhead{Recovered fraction} \\
 & [mJy] & [mJy] & -- }
 \startdata
 ALESS 3.1 & 8.3$\pm$0.4 & 8.7$\pm$0.2 & 1.05$\pm$0.06\\
 ALESS 9.1 & 8.8$\pm$0.5 & 9.1$\pm$0.2 & 1.03$\pm$0.06\\
 ALESS 15.1 & 9.0$\pm$0.4 & 9.6$\pm$0.2 & 1.06$\pm$0.05\\
 ALESS 17.1 & 8.4$\pm$0.5 & 8.8$\pm$0.2 & 1.04$\pm$0.06\\
 ALESS 76.1 & 6.4$\pm$0.6 & 5.0$\pm$0.1 & 0.78$\pm$0.07\\
 ALESS 112.1 & 7.6$\pm$0.5 & 6.1$\pm$0.2 & 0.80$\pm$0.06
 \enddata
 \label{tab:fluxcomp}
 \end{deluxetable*}

%%%%%%%%%%%%%%%%%%%%%%%%%%%%%%%%%%%%%%%%%%%%%%%
\subsection{HST Imaging }
\label{HSTimaging}

\begin{figure*}
\centering
\includegraphics[scale=0.6,trim={0 1cm 0 0cm},clip]{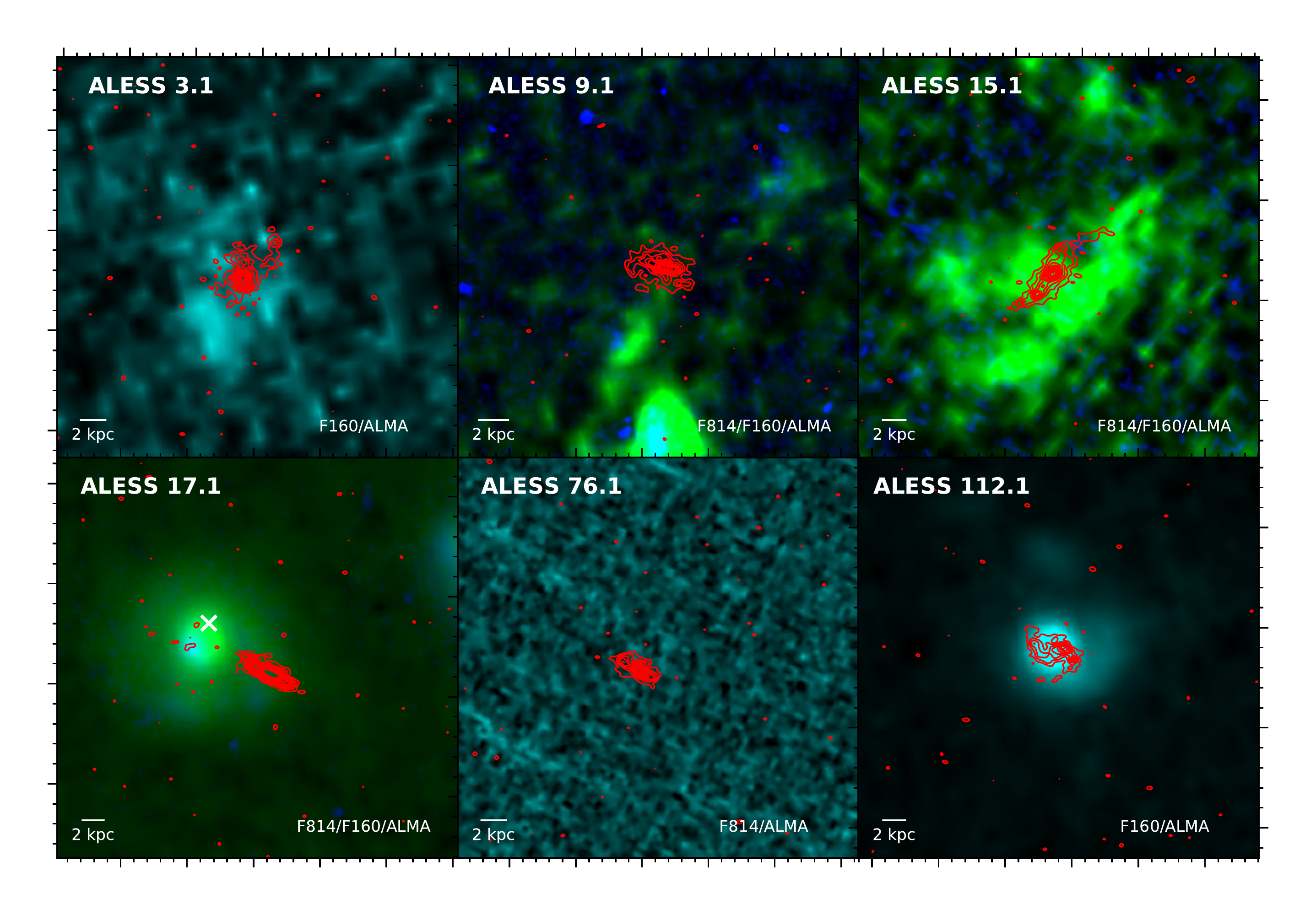}
\caption{4$''$$\times$4$''$ false-color images of the \textit{HST} and ALMA data for each of our sources. Shown are the \textit{HST} $H_{\rm 160}$-band (green), the \textit{HST} $I_{\rm 814}$-band (blue), and the ALMA 870\,$\mu$m emission at 0.08$''$$\times$0.06$''$ resolution (middle column of Figure~\ref{fig:ALESScyc4_ALL}; red contours). The position of the X-ray source in ALESS 17.1 is indicated by the white cross. ALMA contours are shown at 3,6,9,12...$\sigma$, and the \textit{HST} stretch has been adjusted to enhance the visibility of faint emission as needed. This comparison suggests that the ALMA imaging may be revealing the starbursting cores of more extended highly-obscured systems.} 
\label{fig:HSTcomp}
\end{figure*}

We include in our analysis \textit{HST} imaging from the Cosmic Assembly Near-infrared Deep Extragalactic Legacy Survey \citep[CANDELS;][]{Grogin2011, Koekemoer2011} and our own \textit{HST} program \citep{Chen2015}. As presented in \citet{Chen2015}, the combined dataset on all 60 ALESS SMGs covered by these programs has a median point-source sensitivity in the $H_{\rm 160}$-band of $\sim$27.8 mag, corresponding to a 1$\sigma$ depth of $\mu_{\rm H}$ $\sim$ 26 mag arcsec$^{-2}$. The astrometry was corrected on a field-by-field basis using \textit{Gaia} DR1 observations \citep{Brown2016, Prusti2016}. The newly derived solutions were within $\lsim$0.1$''$ in both right ascension and declination from the astrometric solutions previously derived by \citet{Chen2015} from a comparison with the 3.6\,$\mu$m \textit{Spitzer} imaging.

%%%%%%%%%%%%%%%%%%%%%%%%%%%%%%%%%%%%%%%%%%%%%%%%%%%%%%%%%%%%%%%%%%%%%%%%%%%%%%
\section{RESULTS}
\label{results}

Figure~\ref{fig:ALESScyc4_ALL} shows the ALMA maps of our six targeted SMGs, each imaged at three different spatial resolutions. At the redshifts of our targets (Table~\ref{tab:physprops}), 870\,$\mu$m corresponds to a rest-frame wavelength of $\sim$250\,$\mu$m (ranging from 150--350\,$\mu$m), and a beam size of 0.07$''$ corresponds to a typical spatial resolution of $\sim$500\,pc (ranging from 450--600\,pc). All six sources show clear structure on these scales. The significance (both statistically and physically) of these structures will be discussed in more detail in \S\ref{structure}. Before we attempt to interpret the meaning of the observed ALMA structure, we first examine the global ALMA$+$\textit{HST} morphologies of the sources.

\begin{figure*}
\centering
\includegraphics[scale=0.78,trim={0 2.3cm 0 1cm},clip]{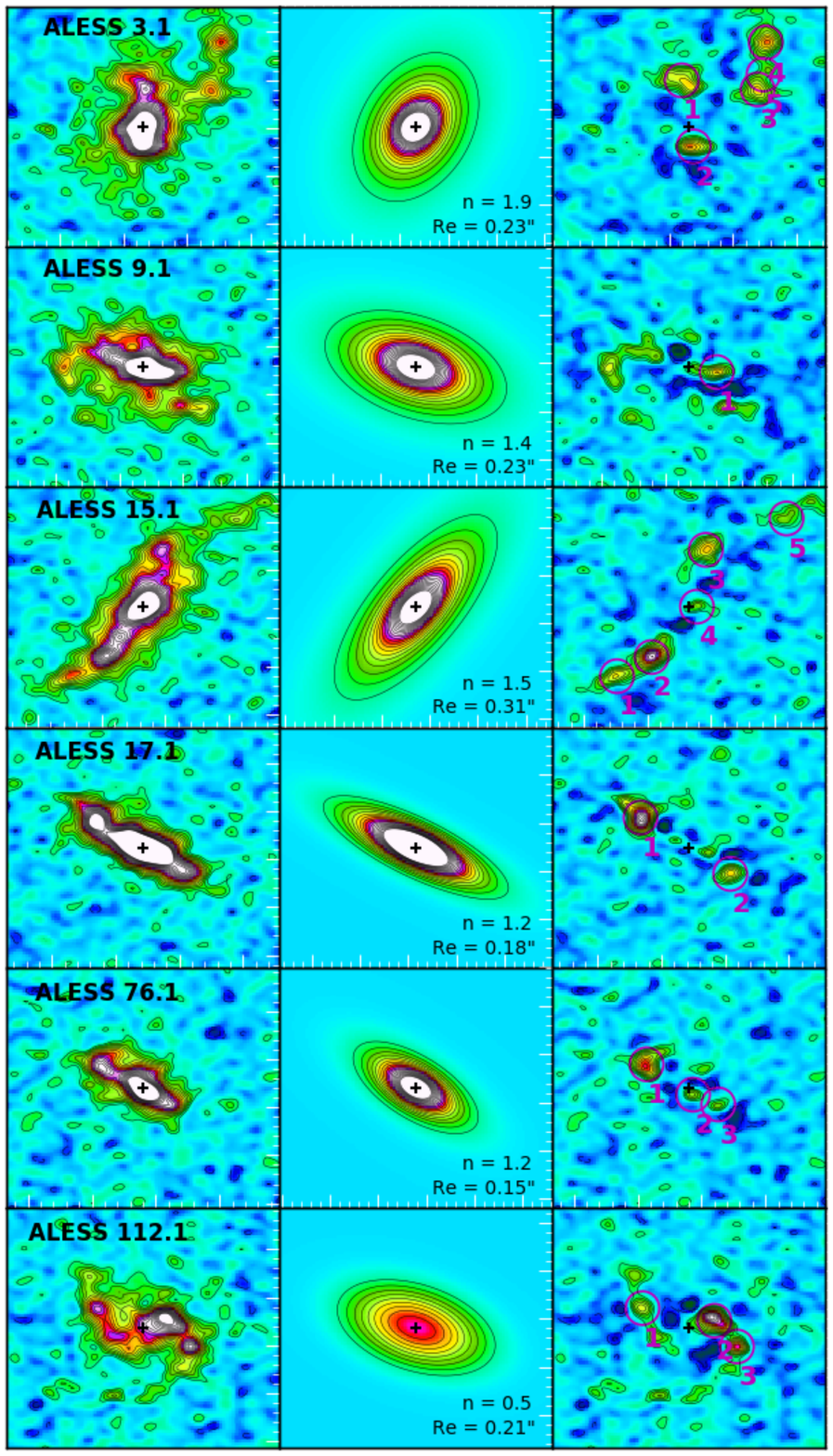}
\caption{{\sc galfit} modeling and substructure identification in our six galaxies as discussed in \S\ref{structure}. Panels (1.0$''$$\times$1.0$''$) show the observed maps with Briggs ($R = +0.5$) weighting and $\sim$0.07$''$/500\,pc resolution (left column); the best-fit S\'ersic profile after masking residual pixels $>$5$\sigma$ iteratively (middle column); and the residual maps resulting from the iterative masking (right column). The black cross marks the center of the model. Contours start at $\pm$2$\sigma$ and go in steps of 1$\sigma$, stopping at 20$\sigma$ for clarity. Structures more significant than the largest negative peak in each map are circled and labeled according to Table~\ref{tab:clumpprops}. All six of the sources studied here show significantly detected complex dusty structure, including evidence for pairs of clump-like structures bracketing the elongated nuclear regions along the major axes of the most inclined sources. We discuss the possibility that we are observing inclined bar$+$ring morphologies in Section~\ref{morphology}. } 
\label{fig:GALFIT}
\end{figure*}

\subsection{\textit{HST} comparison}
\label{HSTcomp}

Figure~\ref{fig:HSTcomp} shows false-color images for our sources constructed using a combination of the ALMA and deep \textit{HST} imaging in one or more bands (\S\ref{HSTimaging}), where the latter allows us to probe the existing unobscured stellar distribution at lower (0.15$''$/1.2\,kpc at $z\sim2.5$) resolution. The first thing to notice is that there is no correlation between the potential clumpy structure revealed in the new ALMA imaging and the \textit{HST} imaging for any of the galaxies. This is because the dust emission traced by ALMA is more compact than the \textit{HST} sources, as noted in previous studies \citep{Simpson2015b, Hodge2016, Chen2017, CalistroRivera2018}. Nevertheless, a careful look at the position of the ALMA emission relative to the rest-frame optical/UV emission can provide insight on these sources. Detailed notes on individual sources follow below. 

\textbf{ALESS 3.1 ($z_{\rm spec} = 3.374$):} The deep $H_{\rm 160}$-band imaging of this source was previously analyzed by \citet{Chen2015}, who reported a single $H_{\rm 160}$-band component with an effective radius of r$_e$ = 5.5$\pm$0.7\,kpc (the S\'ersic index was fixed at $n$=1.0 due to the low S/N of the source). Comparing to our ALMA data, the centroid of this $H_{\rm 160}$-band `component' lies $\sim$0.5$''$ ($\sim$3.5\,kpc) south of the ALMA source, which itself appears embedded in more extended, low-S/N $H_{\rm 160}$-band emission. If the dusty starburst detected by ALMA is centered on the center of mass of this system, then this source may be experiencing significant differential obscuration. 

\textbf{ALESS 9.1 ($z_{\rm spec} = 4.867$):} The \textit{HST} images are blank at the position of the ALMA-detected emission. There is a possible faint detection in the $H_{\rm 160}$-band emission, but it is offset $\sim$0.8$''$ south of the ALMA source. The $I_{\rm 814}$-band CANDELS image is marred by an artifact near the ALMA source position but is otherwise blank.

\textbf{ALESS 15.1 ($z_{\rm phot} = 2.67$):} The source is undetected in the $I_{\rm 814}$-band and has an extended, clumpy morphology in the $H_{\rm 160}$-band imaging. 
Like ALESS 3.1, it is possible that the ALMA emission (which shows a distinct curvature over its $\sim$10\,kpc extent -- see also Figure~\ref{fig:ALESScyc4_ALL}) is centered on a more extended system which is suffering from differential dust obscuration.
 
\textbf{ALESS 17.1 ($z_{\rm spec} = 1.539$):} The false-color image for ALESS 17.1 shows that the bulk of the ALMA 870\,$\mu$m emission lies offset ($\sim$0.75$''$) from a disk galaxy in the $\textit{HST}$ imaging (though we do detect some very faint 870\,$\mu$m emission near the optical galaxy's nucleus). The galaxy detected in ALMA emission is undetected in \textit{HST} imaging. Recent SINFONI imaging of the field (PI Swinbank)
reveals H$\alpha$ emission from both the optically detected galaxy and the ALMA source, indicating that they lie at the same redshift and are therefore likely interacting. Interestingly, this system is also associated with an X-ray AGN \citep{Wang2013}. The position of the X-ray source is indicated in Figure~\ref{fig:HSTcomp} as that reported by \citet{Luo2017}, with an additional astrometric correction for the median offset reported between that X-ray catalog and \textit{Gaia} DR1.  

\textbf{ALESS 76.1 ($z_{\rm spec} = 3.389$):} This source appears completely undetected in the \textit{HST} imaging ($I_{\rm 814}$-band). We note that longer-wavelength ($H_{\rm 160}$-band) imaging is not available.

\textbf{ALESS 112.1 ($z_{\rm spec} = 2.315$):} The ALMA-detected 870\,$\mu$m continuum emission (which shows a prominent curvature over its $\sim$5\,kpc extent) appears by-eye to be colocated with a bright counterpart in the \textit{HST} $H_{\rm 160}$-band imaging. 
The best-fit model to the $H_{\rm 160}$-band imaging has a S\'ersic index of $n$=3.4$\pm$1.3 and an effective radius of r$_e$ = 0.59$''$$\pm$0.05$''$, corresponding to 4.9$\pm$0.4\,kpc. This supports the conclusion that the high surface brightness 870\,$\mu$m (rest-frame FIR) emission is confined to the nucleus of a more extended stellar distribution. 

In summary, despite the depth of the \textit{HST} imaging (\S\ref{HSTimaging}), the stellar emission from a number of the sources is extremely faint or invisible, making it challenging to characterize the rest-frame optical/UV morphologies of the systems. A superficial analysis shows that the majority of the \textit{HST}-detected sources show offsets between the ALMA 870\,$\mu$m emission (tracing the rest-frame FIR) and the peak of the significantly-detected emission in the deep \textit{HST} imaging (tracing the unobscured rest UV/optical emission of these galaxies). These offsets are significant with respect to the relative astrometric accuracy of the two datasets (see \S\ref{HSTimaging}). However, for at least half (3/6) of the sources (and the majority detected in the \textit{HST} imaging), extended \textit{HST} emission surrounds the ALMA emission, indicating that the ALMA imaging may be revealing the heavily-obscured starbursting cores of larger-scale systems. The comparison here highlights the need for sensitive high-resolution, near-/mid-IR imaging of these dusty targets with a telescope such as the upcoming \textit{James Webb Space Telescope} (\textit{JWST}). 
We now turn to the statistical significance and possible interpretations of the new sub-kpc dusty structure revealed by our ALMA data. 

\subsection{Sub-kpc FIR structure}
\label{structure}

The high-resolution ($\sim$500\,pc) images of our six SMGs presented in Figure~\ref{fig:ALESScyc4_ALL} 
are generally dominated by an extended disk-like morphology -- confirming the results of \citet{Hodge2016} based on shallower, lower-resolution data -- but the new high-fidelity data presented here reveal new structures within these disks. We note that all visible structures were evident also in the dirty maps, indicating that they are not artifacts of the \textsc{clean}ing process. 

To assess the significance of the clumpy structure, we fit the galaxies with two-dimensional S\'ersic profiles in {\sc galfit} \citep{Peng2002, Peng2010}, masking residual pixels $>$5$\sigma$ iteratively until the masks converged. This technique ensures that any real positive structure in the disks would not artificially boost the fits of the underlying smooth profiles, resulting in large negative troughs in the residual images. The resulting fits have half-light radii consistent with, and S\'ersic indices that are on-average slightly higher than, those derived without the masking procedure or from the lower-resolution data in \citet{Hodge2016} (with the notable exception of ALESS 112.1, which will be discussed further below). The results of this iterative procedure are shown in Figure~\ref{fig:GALFIT}, in which candidate structures are identified as structures more significant than the largest negative peak in each residual (i.e., S\'ersic-subtracted) image. In general, between 1--5 residual structures are identified in each source at peak SNRs ranging from $\sim$4--15$\sigma$. Some of these structures lie near/within the nuclei and may be unresolved along one or both axes, indicating either real compact structure or a poor-fitting larger-scale profile (e.g., structure \#4 in ALESS 15.1), while others are clear `clumps' in the disk (e.g., structure \#1 in ALESS 17.1). Based on two-dimensional Gaussian fits in the image plane, these structures individually make up a few percent ($\sim$1--8\%) of the total continuum emission from the galaxies, with a combined contribution of $\sim$2--20\% for a given galaxy.  estimate a median systematic error of $\sim$20\% due to remaining structure in the residual maps.
The deconvolved major axes of the structures range from 600\,pc to 1.1\,kpc for the roughly half that are resolved. Their properties are summarized in Table~\ref{tab:clumpprops}.

Even with these high-resolution, high-S/N data, the disk-like component still dominates the emission in these galaxies. The S\'ersic indices we derive for the extended component from the iterative masking and fitting procedure are typically disk-like ($<n>$=1.3$\pm$0.3), consistent with those derived from the lower resolution (0.16$''$) data of a larger sample in \citet{Hodge2016} ($<n>$=0.9$\pm$0.2). One source (ALESS 112.1) has a very low ($n=0.5$) S\'ersic index. This source also has a large clump-like structure identified very near to the nucleus itself, indicating that a S\'ersic profile may not be appropriate for the complex morphology seen here, which also shows a pronounced curvature.  

Beyond the presence of these clumpy structures, their orientation may provide some clue as to their nature. In particular, in at least three of the sources (ALESS 15.1, 17.1, and 76.1), we see a significant clump-like structure on either end of an elongated nuclear region, and oriented approximately along the major axis. We will discuss a possible interpretation for these features in \S\ref{morphology}.

\begin{deluxetable*}{ l c c c c c c c c c c }
\tabletypesize{\small}
%\tablewidth{16cm}
\tablecaption{S\'ersic profile parameters \& properties of the dusty substructures}
\tablehead{
\colhead{Source} & \colhead{R$_{e}$$^{a}$} & \colhead{$n$$^{a}$} & \colhead{$b/a$$^{a}$} & \colhead{Structure$^{b}$} & \colhead{SNR$_{\rm pk}$$^{c}$} & \colhead{S$_{\rm pk}$$^{c}$} & \colhead{S$_{\rm int}$$^{c}$} & \colhead{$f_{\rm flux}$$^{d}$} & \colhead{bmaj$^{e}$} & \colhead{bmin$^{e}$}  \\
& ($''$) & & & & & ($\mu$Jy beam$^{-1}$) & ($\mu$Jy) & (\%) & (pc) & (pc)}
 \startdata
 ALESS 3.1 & 0.23$\pm$0.01 & 1.9$\pm$0.1 & 0.68$\pm$0.02 & 1 & 8.1 & 180$\pm$30 & 530$\pm$100  & 6$\pm$1 & 1100$\pm$300 & 500$\pm$200 \\ 
 			 & & & & 2 & 10.0 & 220$\pm$20 & 250$\pm$40 & 2.6$\pm$0.3 & -- & -- \\
 			 & & & & 3 & 7.3   & 160$\pm$20 & 390$\pm$80 & 5$\pm$1 & 800$\pm$200 & 500$\pm$300 \\ 
 			 & & & & 4 & 8.8   & 190$\pm$20 & 430$\pm$50 & 5.2$\pm$0.7 & 800$\pm$100 & 300$\pm$200 \\
 			 & & & & 5 & 4.2   & 90$\pm$10   & 220$\pm$40 & 2.7$\pm$0.5 & 1100$\pm$200 & 100$\pm$200 \\
 \hline
 ALESS 9.1 & 0.23$\pm$0.01 & 1.4$\pm$0.1 & 0.53$\pm$0.02 & 1 & 8.6   & 190$\pm$20 & 170$\pm$40 & 2.2$\pm$0.3 & -- & -- \\
 \hline
 ALESS 15.1 & 0.31$\pm$0.01 & 1.5$\pm$0.1 & 0.37$\pm$0.02 & 1 & 7.1 & 160$\pm$10 & 340$\pm$40 & 1.7$\pm$0.2 & -- & -- \\
 			    & & & & 2 & 13.0 & 290$\pm$20 & 550$\pm$60 & 6.1$\pm$0.7 & 900$\pm$100 & 200$\pm$100 \\
			    & & & & 3 & 8.6   & 190$\pm$20 & 360$\pm$50 & 4.0$\pm$0.6 & 800$\pm$200 & 300$\pm$100 \\
			    & & & & 4 & 5.2   & 114$\pm$7 & 80$\pm$10 & 1.3$\pm$0.1 & -- & -- \\
			    & & & & 5 & 4.8   & 105$\pm$7 & 200$\pm$20 & 2.2$\pm$0.2 & 900$\pm$100 & 270$\pm$70 \\
 \hline
 ALESS 17.1 & 0.18$\pm$0.01 & 1.2$\pm$0.1 & 0.26$\pm$0.01 & 1 & 15.5 & 340$\pm$40  & 660$\pm$100 & 8$\pm$1 & 800$\pm$100 & 200$\pm$200 \\
 			   & & & & 2 & 8.2 & 180$\pm$10 & 180$\pm$20 & 2.1$\pm$0.2 & -- & -- \\
 \hline
 ALESS 76.1 & 0.15$\pm$0.01 & 1.2$\pm$0.1 & 0.40$\pm$0.02 & 1 & 10.3 & 230$\pm$30 & 400$\pm$90 & 6$\pm$2 & 600$\pm$200 & 200$\pm$300 \\
 			   & & & & 2 & 6.4 & 140$\pm$10 & 60$\pm$20 & 2.2$\pm$0.3 & -- & -- \\
 			   & & & & 3 & 5.0  & 110$\pm$20 & 60$\pm$20 & 1.7$\pm$0.3 & -- & -- \\
 \hline
 ALESS 112.1 & 0.21$\pm$0.01 & 0.5$\pm$0.1 & 0.52$\pm$0.04 & 1 & 7.0 & 150$\pm$10 & 170$\pm$20 & 2.0$\pm$0.2 & -- & -- \\
			  & & & & 2 & 16.0 & 350$\pm$50 & 540$\pm$100 & 7$\pm$1 & 600$\pm$200 & 200$\pm$200 \\
			  & & & & 3 & 10.9 & 240$\pm$20 & 210$\pm$40 & 3.2$\pm$0.4 & -- & -- 
 \enddata
 \tablecomments{$^{a}$Parameters from the best-fit S\'ersic profile.\\
 $^{b}$Structure number as labeled in Figure~\ref{fig:GALFIT}.\\
 $^{c}$Peak signal-to-noise, peak flux density, and integrated flux density of the feature from a two-dimensional Gaussian fit in the image plane.\\
 $^{d}$Fraction of the total flux density of the galaxy, measured from the compact configuration (Cycle~0) values given in Table~\ref{tab:fluxcomp}. \\
 $^{e}$Deconvolved sizes. Blank entries indicate the structure is unresolved at the current resolution (0.08$''$$\times$0.06$''$) and sensitivity. \\ }
 \label{tab:clumpprops}
 \end{deluxetable*}

\begin{deluxetable*}{ l l l l }[]
\tabletypesize{\small}
%\tablewidth{10cm}
\tablecaption{Inferred star formation rate densities}
\tablehead{
\colhead{Source ID} & \colhead{Mean $\Sigma_{\mathrm{SFR}}$} & \colhead{Peak $\Sigma_{\mathrm{SFR}}$ at 0.07$''$} & \colhead{Peak $\Sigma_{\mathrm{SFR}}$ at 0.05$''$} \\
 & [M$_{\odot}$ yr$^{-1}$ kpc$^{-2}$] & [M$_{\odot}$ yr$^{-1}$ kpc$^{-2}$] & [M$_{\odot}$ yr$^{-1}$ kpc$^{-2}$] }
 \startdata
 ALESS 3.1 & 33$^{+8}_{-15}$ & 180$^{+31}_{-30}$ & 212$^{+40}_{-39}$\\
 ALESS 9.1 & 102$^{+27}_{-32}$ & 547$^{+102}_{-93}$ & 575$^{+116}_{-108}$\\
 ALESS 15.1 & 7$^{+3}_{-3}$ & 63$^{+26}_{-29}$ & 84$^{+35}_{-39}$\\
 ALESS 17.1 & 13$^{+3}_{-3}$ & 66$^{+5}_{-6}$ & 77$^{+6}_{-8}$\\
 ALESS 76.1 & 44$^{+15}_{-26}$ & 129$^{+39}_{-35}$ & 163$^{+51}_{-45}$\\
 ALESS 112.1 & 13$^{+3}_{-4}$ & 45$^{+9}_{-9}$ & 55$^{+12}_{-12}$
 \enddata
 \label{tab:SFRD}
 \end{deluxetable*}

\begin{figure*}
\centering
\includegraphics[scale=0.6,trim={3cm 1.5cm 0cm 2cm},clip]{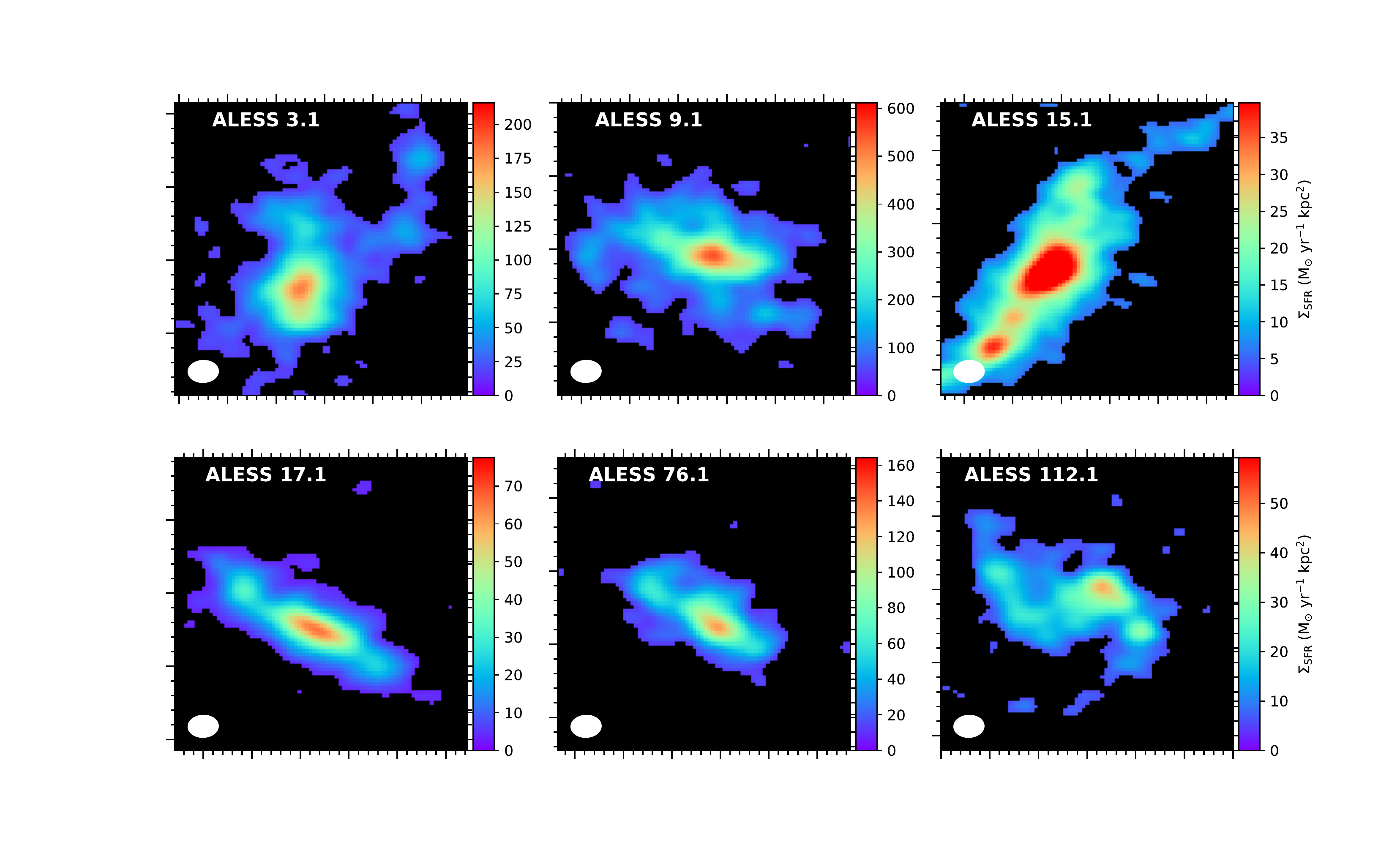}
\caption{SFR surface density ($\Sigma_{\rm SFR}$) maps at $\sim$0.07$''$/500\,pc resolution (corresponding to the middle column of Figure~\ref{fig:ALESScyc4_ALL}), where emission below 3$\sigma$ has been masked. The beam is shown as the white ellipse in the bottom left-hand corner.  
By taking the global SFRs and dust temperatures derived for the galaxies through multi-wavelength SED fitting (Table~\ref{tab:physprops}), we find that the range of $\Sigma_{\rm SFR}$ probed varies between galaxies by over an order of magnitude. This is largely due to the similar S$_{\rm 870}$ values and sizes but very different (global) dust temperatures assumed for the galaxies.} 
\label{fig:SFRD}
\end{figure*}

\subsection{Star formation rate surface density maps}
\label{SFRDmaps}

While the long-wavelength submillimeter emission in high-redshift galaxies can be used to trace the total ISM mass via empirical calibrations \citep[e.g.,][]{Scoville2014, Scoville2016, Scoville2017}, it also correlates with the total star formation rate via the Kennicutt Schmidt star-formation law. For very dust-obscured galaxies like SMGs which are difficult to observe in other commonly-used resolved SFR tracers (e.g., H$\alpha$), studies often rely on high-resolution submillimeter imaging to create maps of resolved star formation rate surface density \citep[$\Sigma_{\rm SFR}$; e.g.,][]{Hodge2015, Hatsukade2015, Chen2017, Canameras2017}. 
This is done by assuming that the variations in the observed submillimeter flux correlate with variations in the local star formation rate, and scaling the total SFR by the observed-to-total ALMA 870\,$\mu$m flux density per beam across a source. The technique relies on having total (global) SFRs for each galaxy which are well-determined through multi-wavelength SED fitting. More critically, it effectively assumes that there are no variations in dust temperature (T$_{\rm d}$) or emissivity index ($\beta$) within the sources, which is unlikely to be correct. Nevertheless, it provides a first estimate of the distribution of $\Sigma_{\rm SFR}$ in these sources on $\sim$500\,pc scales. 

The total far-infrared luminosities (and thus SFRs) for our galaxies are well-constrained by the SEDs for the sources, which have been modified from those presented in \citet{daCunha2015} to include updated redshift information and additional (unresolved) submillimeter observations in ALMA's Band 4 (da Cunha et al.\,in prep). Following the above method, we created maps of star formation rate surface density ($\Sigma_{\rm SFR}$) for our six sources\footnote{In calculating $\Sigma_{\rm SFR}$ in units of M$_{\odot}$ yr$^{-1}$ kpc$^{-2}$ for each beam at our resolution, we note that beam area is defined as $\pi/(4\times\ln(2))\times$b$_{\rm maj}$$\times$b$_{\rm min}$.} (Figure~\ref{fig:SFRD}). 
We show the 0.07$''$-resolution maps in Figure~\ref{fig:SFRD}. The peak values from maps at both resolutions, as well as the galaxy-averaged values calculated using the half-light radii as 0.5$\times$SFR/($\pi$R$_{e}^{2}$), are listed in Table~\ref{tab:SFRD}.

The first thing to notice about Figure~\ref{fig:SFRD} is that the peak $\Sigma_{\rm SFR}$ on $\sim$500\,pc scales varies by over an order of magnitude between galaxies. As the peak 870\,$\mu$m flux densities only vary between galaxies by at most a factor of two, and the physical scale of the emission is similar between galaxies, this is not solely a result of different observed flux density distributions. Rather, this large variation in peak $\Sigma_{\mathrm{SFR}}$ can be traced back to intrinsically different total star formation rates (ranging from $\sim$150--1500 M$_{\odot}$ yr$^{-1}$), and ultimately to different physical conditions (dust luminosities and dust temperatures) in the sources (Table~\ref{tab:physprops}). These different dust temperatures/luminosities are constrained by the peak of the dust SED, which is typically reasonably well-sampled in these sources: all six sources have five photometric data points between $\sim$200\,$\mu$m and $\sim$1.2\,mm (observed frame), with only one source (ALESS 76.1) constrained by upper limits alone in the \textit{Herschel} bands \citep{Swinbank2014}. We also note that this large range of SFRs is not driven by our particular choice of SED-fitting code \citep[\sc{magphys};][]{daCunha2015}, as instead using simple modified blackbody fits with, e.g., the \citet{Kennicutt1998} IR SFR relation (re-scaled to the \citet{Chabrier2003} IMF) returns the same results \citep{Swinbank2014}. Physically, the measurement of a colder integrated dust temperature could indicate a larger contribution from dust heated by older stars \citep{daCunha2008}, or it could indicate that the stellar radiation field seen by dust grains is not as intense. This is partly a selection effect, as the coldest sources are primarily at lower-redshifts. Alternately, it could also be an artifact introduced in the SED modeling by assuming optically-thin dust when it is indeed optically thick, depleting the emission at the shorter infrared wavelengths \citep[e.g.,][]{Scoville2013book, Simpson2017}.

An artifact of the difference in absolute scaling between galaxies is that the faintest $\Sigma_{\rm SFR}$ we are sensitive to also varies between galaxies. For ALESS 9.1 (which has the highest peak $\Sigma_{\rm SFR}$), the 3$\sigma$ cutoff corresponds to 50 M$_{\odot}$ yr$^{-1}$ kpc$^{-2}$. In ALESS 15.1, on the other hand, the 3$\sigma$ cutoff corresponds to 2.6 M$_{\odot}$ yr$^{-1}$ kpc$^{-2}$. This limit is (again) affected by the assumption of a single (global) temperature over the sources.

Another assumption in the above analysis is that the rest-frame FIR emission is due to star formation rather than AGN activity. While this is generally thought to be true for the SMG population \citep[e.g.,][]{Alexander2005,Laird2010}, we note that one of our sources (ALESS 17.1; $L_{\rm 0.5-8keV,corr}$ $=$ 1.2 $\times$ 10$^{43}$\,ergs s$^{-1}$) was classified by \citet{Wang2013} as an AGN based on its low effective photon index ($\Gamma_{\rm eff} < 1$), indicating a hard X-ray spectrum of an absorbed AGN.  
Due to its low $L_{\rm 0.5-8 keV,corr}$/$L_{\rm FIR}$ ratio, however, \citet{Wang2013} concluded that it almost certainly had little to no AGN contribution in the FIR band. Indeed, it is interesting to note that the peak $\Sigma_{\rm SFR}$ of ALESS 17.1 ($\sim$75 M$_{\odot}$ yr$^{-1}$ kpc$^{-2}$) is actually on the lower side of the range for the sources studied in this work, perhaps indicating that the AGN is not even dominant on the scales ($\sim$500\,pc) probed here.

\begin{figure}
\centering
\includegraphics[scale=0.57,trim={0 0 0cm 0cm},clip]{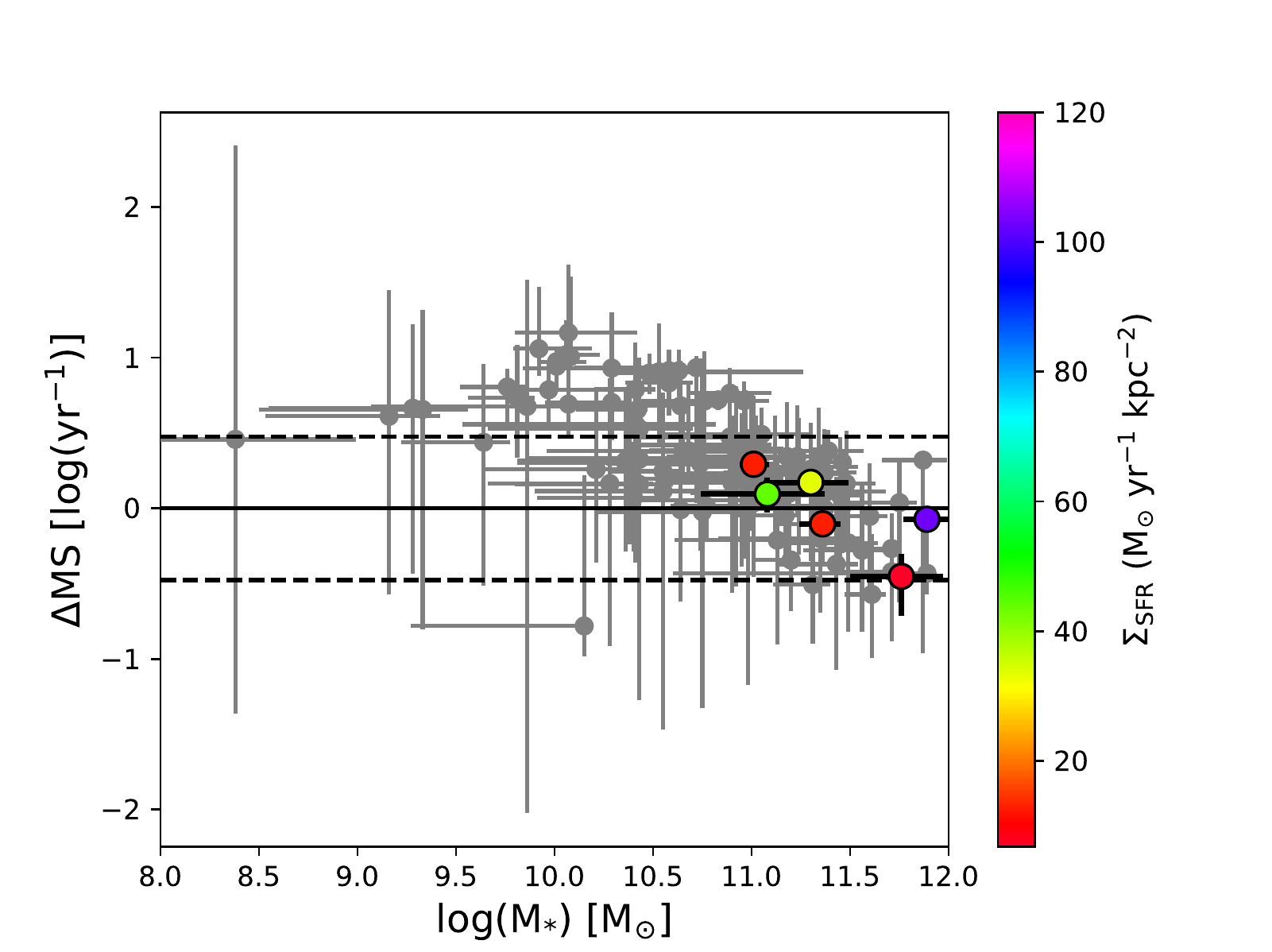}
\caption{Distance from the star-forming SFR--mass trend ($\Delta$MS$=$SFR/SFR$_{\rm MS}$) versus stellar mass for the galaxies studied in this work, where the data points are color-coded by galaxy-averaged SFR surface density. The gray points show the full ALESS SMG sample from \citet{daCunha2015}. As in \citep{daCunha2015}, the definition of the SFR--mass trend (solid line) is from \citet{Speagle2014}, and the dashed lines indicate a factor of three above/below this relation. The error bars on the full ALESS sample are larger as they include a marginalization over the redshift, which was a fitted parameter in \citet{daCunha2015}. Keeping in mind the considerable uncertainties in the creation of such a plot, we see that the six galaxies studied in this work are consistent with the SFR--mass trend for massive galaxies at their redshifts, and there is no correlation with total $\Sigma_{\rm SFR}$ within the sample.}
\label{fig:MScomp}
\end{figure}

\subsection{Relation to the SFR--mass trend}
\label{MScomp}

There has been significant discussion in the recent literature about the relation of SMGs to the SFR--mass trend \citep[e.g.,][]{Noeske2007, Daddi2007a}. In particular, some studies find that SMGs are (on average) offset above the SFR--mass trend in the `starburst regime' \citep[e.g.,][]{Danielson2017}, while others argue that the majority are consistent with the high-mass end of the relation \citep[e.g.,][]{Koprowski2016}. In their study of the full sample of ALESS SMGs, \citet{daCunha2015} found that $\sim$50\% of $z\sim2$ SMGs are consistent with lying on the SFR--mass trend, and that this fraction increases at higher redshift, where the trend evolves to higher values of SFR.

There are significant uncertainties involved in placing any one SMG on this trend, as systematic uncertainties on the stellar mass, star formation rate, and definition of the SFR--mass trend itself \citep[e.g.,][]{Whitaker2012, Whitaker2014, Speagle2014, Tomczak2016} can easily shift the points by an order of magnitude along a given axis.
In particular, there is considerable uncertainty in deriving robust stellar masses for these extremely dusty sources \citep[e.g.,][]{Hainline2011, Michalowski2014, daCunha2015} -- a difficulty which is highlighted by the \textit{HST} non-detections seen in Figure~\ref{fig:HSTcomp}. For these sources, the stellar masses are constrained mainly through detections in the IRAC bands and may carry significant systematic uncertainties (Figure~\ref{fig:MScomp}). In addition, there is also considerable uncertainty in the definition of the SFR--mass trend itself \citep[e.g.,][]{Whitaker2012, Speagle2014}. Nevertheless, it is interesting to consider where the galaxies targeted in this work fall with respect to the SFR--mass trend and the overall population of ALESS galaxies, particularly as they constitute some of the brightest submillimeter galaxies in the sample \citep{Hodge2013a} and yet have values of peak $\Sigma_{\rm SFR}$ which vary by over an order of magnitude (\S\ref{SFRDmaps}). In Figure~\ref{fig:MScomp}, we show the positions of the galaxies studied in this work in relation to the properties of the full ALESS sample as derived in \citet{daCunha2015}. All six of our galaxies are consistent with the SFR--mass trend for massive galaxies at their redshifts, and thus we find no immediate evidence for a correlation between position with respect to the SFR--mass trend and $\Sigma_{\rm SFR}$.

\begin{figure*}
\centering
\includegraphics[scale=0.75,trim={0 0 0 0cm},clip]{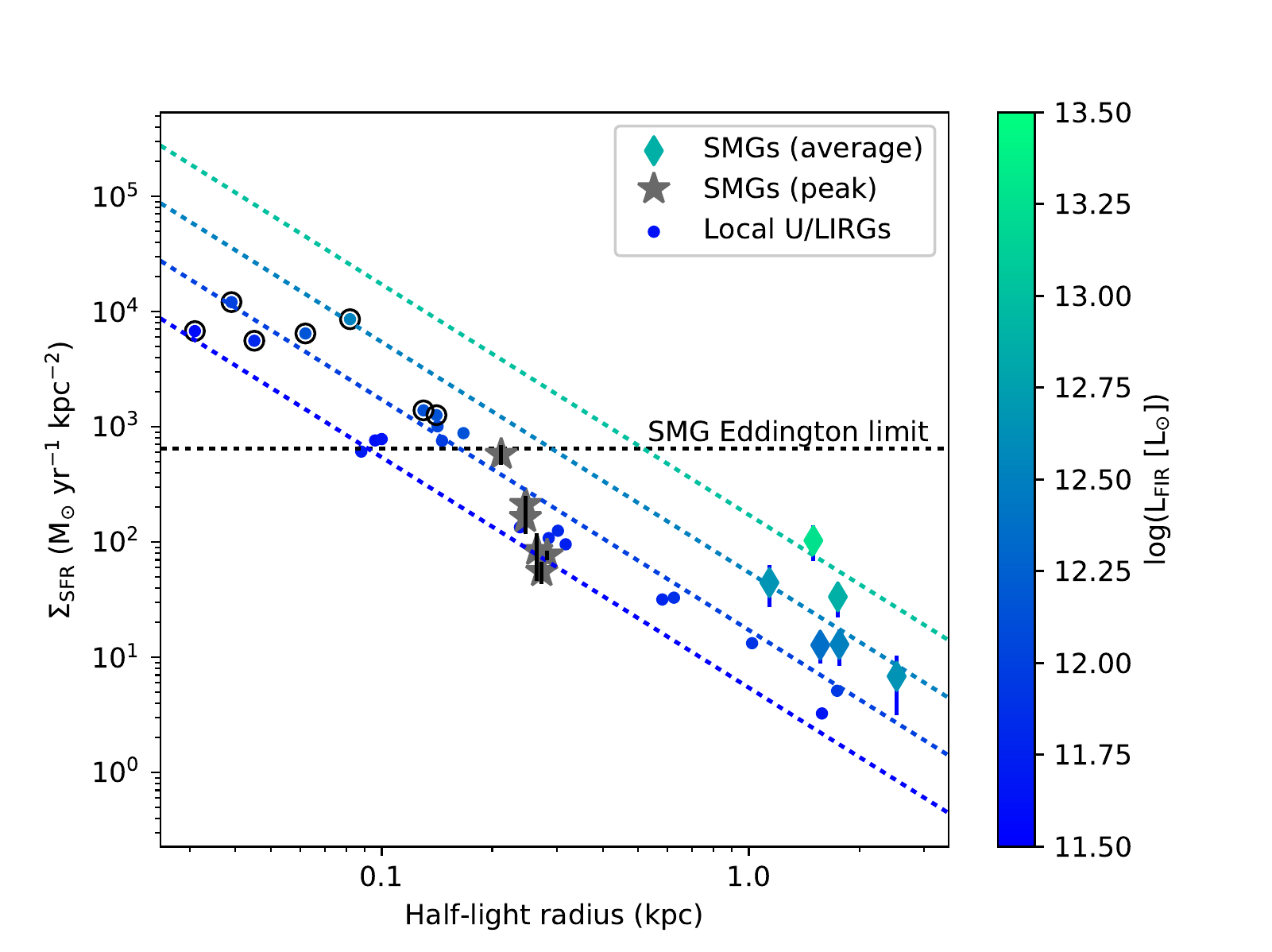}
\caption{Star formation rate surface density ($\Sigma_{\mathrm{SFR}}$) versus half-light radius for local U/LIRGs and the SMGs studied in this work. The local U/LIRGs come from \citet{BarcosMunoz2017}, where the $\Sigma_{\mathrm{SFR}}$ values are galaxy-averaged (rescaled to the Kennicutt (1998) IR-SFR conversion with the \citet{Chabrier2003} IMF) and the half-light radii are the equivalent circular radii of the sources as observed at 33 GHz. For the SMGs in this work, both the galaxy-averaged and peak $\Sigma_{\mathrm{SFR}}$ values are shown, where the latter are calculated at our highest resolution (equivalent to half-light radii of $\sim$250\,pc, with slight variations due to redshift). Both the local U/LIRGs and galaxy-averaged SMG points are color-coded by total FIR luminosity of the galaxy. Dashed diagonal lines indicate lines of constant FIR luminosity assuming the \citet{Kennicutt1998} SFR$_{\rm IR}$ calibration (re-scaled to the \citet{Chabrier2003} IMF). The horizontal dashed line indicates the estimated Eddington-limited SFR density for these SMGs assuming the optically thick limit in a warm starburst (\S\ref{SFintensity}). While \citet{BarcosMunoz2017} find that seven sources have IR surface brightnesses which exceed the characteristic limit for radiation-pressure on dust for optically-thick U/LIRGs (circled points; see Section~\ref{SFintensity}), none of the SMGs exceed the Eddington limit on the resolved scales probed here.  }
\label{fig:SFRDcomp}
\end{figure*}

%%%%%%%%%%%%%%%%%%%%%%%%%%
\section{Discussion}
\label{discussion}

\subsection{The intensity of the star formation}
\label{SFintensity}

Locally, radiation pressure on dust is thought to play an important role in regulating the star formation in the dense, optically thick centers of ULIRGs \citep[e.g.,][]{Scoville2003, Thompson2005}.
Recent ALMA work on SMG sizes has already demonstrated that they lie well below this `Eddington limit' on galaxy-wide scales \citep[e.g.,][]{Simpson2015b}. 
We now investigate whether the SMGs also lie below the Eddington limit on the small physical scales probed here. To do this, we note that the theoretical value of the Eddington limit depends on the assumed physical conditions of the source(s). In particular, one important consideration is whether the galaxies are assumed to be optically thick to the re-radiated FIR emission (i.e., $\tau_{\rm FIR}$ $\gtrsim$ 1) in the dense centers of the starbursts. 
We can estimate whether this is true for our sources by solving the equation: 
\begin{equation}
I_{\nu,obs} = B_{\nu,rest} (1-e^{-\tau_{\nu,rest}}) 
\end{equation}
where $\tau_{\nu}$ is the optical depth at frequency $\nu$, $B_{\nu,rest}$ $=$ $B_{\nu,rest}$(T) is the Planck function, and $I_{\nu,obs}$ is the specific intensity. The specific intensity can be calculated from the observations as 
\begin{equation}
I_{\nu,obs} = 0.5 S_{\nu,obs} (1+z)^3 / \Omega_{\nu,obs}
\end{equation}
where $S_{\nu,obs}$ is the observed (peak) flux density at our highest resolution, $z$ is the redshift, and $\Omega_{\nu,obs}$ 
is the synthesized beam solid angle defined as $\pi/(4\times\ln(2))\times$b$_{\rm maj}$$\times$b$_{\rm min}$. Here we approximate the temperature T required by the Planck function as the (global) dust temperature measured from the SEDs by {\sc MAGPHYS} (Table~\ref{tab:physprops}), but we note that 
a more robust calculation of the true optical depths would require multi-frequency, resolved data in the rest-frame far-infrared to break the degeneracy between dust temperature and optical depth \citep[e.g.,][]{Simpson2017} as well as to map the temperature gradient within the sources. Nevertheless, solving for $\tau_{\nu,rest}$ using the above equations results in values of $\tau$ $>$ 1 for rest-frame wavelengths of $\sim$150-350\,$\mu$m, and the same is true if we assume the central temperature is a factor of a $\sim$few higher. This result is consistent with that expected based on the typical gas surface densities in these sources \citep{Simpson2015b} for the `warm starburst' regime \citep{Andrews2011}. We therefore assume that the dense centers of our sources are optically thick to FIR photons.

In this optically thick limit for warm starbursts, the Eddington flux is then shown by \citet{Andrews2011} to be 
\begin{equation}
F_{\mathrm{Edd}} \sim 10^{13} L_{\odot} \mathrm{kpc}^{-2} f_{\mathrm{gas}}^{-1/2} f_{\mathrm{dg, 150}}^{-1}
\end{equation}
where $f_{\mathrm{gas}}$ is the gas mass fraction. Noting that most of the bolometric luminosity from SMGs is emitted in the IR, and using the IR-based SFR calibration of 
\citet{Kennicutt1998} (re-scaled to a \citet{Chabrier2003} IMF), 
we convert this to an Eddington-limited SFR density of 
\begin{equation}
(\Sigma_{\mathrm{SFR}})_{\mathrm{Edd}} \sim 7.2 M_{\odot} \mathrm{yr}^{-1} \mathrm{kpc}^{-2} f_{\mathrm{gas}}^{-1/2} f_{\mathrm{dg}}^{-1}
\end{equation}
where we note that $f_{\rm dg, 150}$ $=$ $f_{\rm dg}$ $\times$ 150.
Assuming the same dust-to-gas ratio as above (1/90) and adopting a gas fraction of unity as the most extreme scenario, we derive a lower limit on the Eddington-limited $\Sigma_{\mathrm{SFR}}$ of $\sim$650 M$_{\odot}$ yr$^{-1}$ kpc$^{-2}$. As seen in Figure~\ref{fig:SFRDcomp}, none of the SMGs exceed this limit, even on the resolved scales probed here, and even in the individual clump-like structures (with the caveats stated above).  

In Figure~\ref{fig:SFRDcomp}, we also compare our galaxy-averaged and peak $\Sigma_{\mathrm{SFR}}$ values with the galaxy-averaged $\Sigma_{\mathrm{SFR}}$ values derived for 22 local luminous and ultra-luminous galaxies (U/LIRGs) from \citet{BarcosMunoz2017}. These U/LIRGs were selected from the \textit{IRAS} Revised Bright Galaxy Sample \citep[RBGS;][]{Sanders2003} as 22 of the most luminous sources in the northern sky, and they have a median FIR luminosity of $\sim$10$^{11.8}$ L$_{\odot}$, corresponding to a median SFR of $\sim$80 M$_{\odot}$ yr$^{-1}$. 
Their $\Sigma_{\mathrm{SFR}}$ values were calculated using IR-based SFRs and assuming that the 33 GHz size reflects the distribution of the star formation. Their physical resolution ranged from 30--720\,pc, and in some cases, the sources were only marginally resolved. 

We see in Figure~\ref{fig:SFRDcomp} that the average $\Sigma_{\mathrm{SFR}}$ values and half-light radii for the local U/LIRGs are fairly tightly correlated. The scatter in the correlation can be attributed to the range in total FIR luminosities for the U/LIRGs. The local U/LIRGs also span a much wider range in galaxy-averaged $\Sigma_{\mathrm{SFR}}$ than the SMGs, which is largely due to the fact that the physical sizes of the U/LIRGs span $>$1 dex, whereas the SMG sizes are fairly homogenous and much larger on average\footnote{Note that no pre-selection was made in our sample on morphology or scale of the submillimeter emission, as discussed in \S\ref{obs}.}. For a given total source size, however, the SMGs can have galaxy-averaged $\Sigma_{\mathrm{SFR}}$ values of up to an order of magnitude higher than U/LIRGs. This can be attributed to the larger total FIR luminosities of the SMGs. Looking at this from a different perspective, the results in Figure~\ref{fig:SFRDcomp} suggest that SMGs are able to sustain a given galaxy-averaged $\Sigma_{\rm SFR}$ over much larger physical extents.

However, we also see from Figure~\ref{fig:SFRDcomp} that local U/LIRGs can achieve even higher $\Sigma_{\mathrm{SFR}}$ values on smaller physical scales than are observable here. Indeed, \citet{BarcosMunoz2017} found that seven of their sources have IR surface brightnesses which exceed the characteristic limit of $\sim$10$^{13}$ L$_{\odot}$ kpc$^{-2}$ for radiation-pressure on dust in optically-thick U/LIRGs -- even averaging over the sources -- {which \citet[][]{Thompson2005} argue is consistent with that expected for a central gas fraction of $f_{\rm g}$ $\sim$ 0.5. For an even} more conservative gas fraction of 0.3, five of the U/LIRGs would still exceed the limit, indicating that they may be Eddington-limited starbursts. The apparent super-Eddington values could then be due to one of the assumptions in the calculation breaking down, such as the assumption of equilibrium in the system through the generation of a galactic wind. Interestingly, the peak $\Sigma_{\mathrm{SFR}}$ values for the SMGs measured at the highest resolution (equivalent to half-light radii of $\sim$250\,pc) are similar to those of U/LIRGs with that same \textit{total} size. As the U/LIRGs only exceed the Eddington limit on smaller physical scales, this could indicate that even higher resolution ($<$500\,pc FWHM) observations would be necessary to observe super-Eddington star formation in SMGs.

One important caveat in the above analysis is the previously stated assumption of a single dust temperature across the sources. This assumption is unlikely to be true based on both detailed studies of resolved local galaxies \citep[e.g.,][]{Pohlen2010, Engelbracht2010, Galametz2012} as well as from radiative transfer modeling of the dust versus CO extents from a stacking analysis of the ALESS SMGs specifically, where the observations are well-fit by radially decreasing temperature gradients \citep{CalistroRivera2018}. Assuming a dust temperature gradient that decreased with radius would change the distribution of the $\Sigma_{\mathrm{SFR}}$, causing it to peak at higher values in the center and decrease more rapidly in the outskirts. Determining the magnitude of this effect will require resolved, high-S/N multi-band continuum mapping of these high-redshift sources to map their internal dust temperature gradients with ALMA.

\subsection{Dusty substructure in SMGs}
\label{substructure}
 
The high-resolution, high-S/N ALMA 870\,$\mu$m imaging of SMGs presented in this work confirms the disk-like morphology of the dusty star formation in these galaxies \citep{Hodge2016, Gullberg2018}, which -- although very compact relative to the \textit{HST} imaging -- is more extended than similarly luminous local galaxies in the FIR (e.g., Arp 220). If we interpret the structures we observe in our galaxies as star-forming `clumps' -- defined as discrete star-forming regions such as those claimed in rest-frame optical/UV imaging \citep[e.g.,][]{Guo2012, Guo2015}, near-infrared integral field spectroscopy \citep[e.g.,][]{ForsterSchreiber2011}, and molecular gas imaging \citep[e.g.,][]{Tacconi2010, Hodge2012} of high-redshift galaxies -- 
 this allows us to place some first constraints on the importance of these structures to the global star formation in these massive, dusty sources. We find that they each contain only a few percent of the emission in a given galaxy, with a combined contribution of $\sim$2--20\% and an additional systematic error of 20\% (\S\ref{structure}). Assuming a constant internal dust temperature (\S\ref{SFRDmaps}), this would imply that kpc-scale clumps are not the dominant sites of star formation in these SMGs. If the clump-like structures we observe trace sites of young massive star formation, then the dust temperature in these regions may be higher than the quiescent regions \citep[e.g.,][]{Galametz2012}, implying that their fractional contribution to the global SFRs may be higher.
We also note that the (luminosity-weighted) temperatures estimated by our global measurements are likely to be dominated by the star-forming regions seen in the continuum maps \citep[e.g.,][]{Utomo2019}.

For comparison, hydro-cosmological zoom simulations of giant clumps in $1<z<4$ disk galaxies have previously examined the contribution of both in-situ (via violent disk instability) and ex-situ (via minor mergers) clumps to the total SFRs, finding that a central `bulge clump' alone usually accounts for 23\% (on average) of the total SFR \citep{Mandelker2014}. This is a larger contribution than we identified in any of our clump-like structures -- regardless of position relative to the bulge -- although we are implicitly assuming that such clumps would still be identifiable in our S\'ersic fits to the continuum emission (as opposed to the molecular gas line emission in 3D, as done in the simulations). Only considering off-center clumps, \citet{Mandelker2014} find an average SFR fraction in clumps of 20\% (range 5--45\%) -- somewhat higher than we estimate, though these clumps are distributed over larger areas and the galaxies themselves are generally less massive ((0.2--3)$\times$10$^{11}$ M$_{\odot}$ yr$^{-1}$) and less highly star-forming than the galaxies imaged here. 

The clumpy structures that we do significantly detect have (deconvolved) sizes ranging from unresolved (at 500\,pc resolution) to $\sim$1\,kpc. Following \citet{Gullberg2018} and estimating the gas surface density from the global $\Sigma_{\rm SFR}$, we find that obtaining a Jeans length comparable to the largest clump-like feature observed in each source would require velocity dispersions of $\gtrsim$60--150 km s$^{-1}$ (Table~\ref{tab:veldisp}).
While we do not have measured velocity dispersions for these sources specifically (though see \S\ref{morphology}), observations of other SMGs (lensed and unlensed) suggest values of 10--100 km s$^{-1}$ \citep{Hodge2012, DeBreuck2014, Swinbank2015}. Taking the value of 40 km s$^{-1}$ measured previously for one source from the full ALESS sample \citep[ALESS 73.1;][]{DeBreuck2014} gives Jeans lengths ranging from 50--400\,pc. Therefore, while the above calculation assumes both velocity dispersion and gas surface density,  
it is possible 
that the largest clump-like structures that we observe may either be blends of smaller structures at the current beam size, or may not be self-gravitating.  
We attempt to place further constraints on the velocity dispersion for a subset of sources below in \S\ref{morphology}. 

\begin{deluxetable}{ l c c }[]
\tabletypesize{\small}
%\tablewidth{10cm}
\tablecaption{Constraints on velocity dispersions}
\tablehead{
\colhead{Source ID} & \colhead{Jeans length constraint$^a$} & \colhead{Toomre constraint$^b$} \\
 & [km s$^{-1}$] & [km s$^{-1}$]
 }
 \startdata
 ALESS 3.1  & $>$150  & -- \\
 ALESS 9.1  & --        & -- \\
 ALESS 15.1 & $>$60    & $<$70 \\
 ALESS 17.1 & $>$75    & $<$90 \\
 ALESS 76.1 & $>$100  & $<$160 \\
 ALESS 112.1 & $>$70  & -- 
 \enddata
 \tablecomments{$^a$Minimum velocity dispersion necessary for the Jeans length to equal the major axis of the largest clump-like structure in that source (Table~\ref{tab:clumpprops}).\\
 $^b$Maximum velocity dispersion allowable for a Toomre stability parameter $Q<1$ in the sources with potential bar-like features (Section~\ref{morphology}).
 }
 \label{tab:veldisp}
 \end{deluxetable}

\subsection{Evidence for interaction, bars, rings, and spiral arms?}
\label{morphology}
	
A comparison between the high-resolution ALMA images and deep \textit{HST} imaging provides further insight into these highly star-forming sources. In particular, for one source (ALESS 17.1), we see a submillimeter component that is significantly (spatially) offset from a separate optically-detected disk galaxy. This offset is now confirmed as significant thanks to the resolution of ALMA and the astrometric solutions of \textit{Gaia}. SINFONI spectroscopic imaging indicates that the submillimeter- and optically-detected galaxies are at the same redshift (M. Swinbank, personal communication), and thus we are likely witnessing an interaction-induced starburst in the ALMA source, which is itself undetected in the optical. 
Interestingly, this is also the only one of our sources associated with an X-ray AGN \citep{Wang2013}. 

A more general observation from the \textit{HST} comparison is that for at least half (3/6) of the sources (and the majority detected in the \textit{HST} imaging), a careful inspection suggests that the ALMA emission may be centered on disturbed and/or partially obscured optical disks. This then suggests that the ALMA imaging in these cases is tracing the dusty cores of more extended systems, and it also aids in the interpretation of the dusty substructure in the global picture. 

In particular, in two of these sources (112.1/15.1), the morphology of the high-fidelity ALMA imaging shows a very clear curvature reminiscent of either spiral arms or the star-forming knots in an interaction/merger such as the Antennae \citep{Klaas2010}. In this case, the scale of the emission is an important clue. From Figure~\ref{fig:HSTcomp}, it is clear that the dusty structure revealed by the ALMA imaging is tracing the inner $\sim$5--10\,kpc of the systems, and is thus inconsistent with larger-scale tidal features.

The curvature seen in the 870\,$\mu$m emission of ALESS 112.1 and 15.1 may then be revealing star-forming spiral structure, potentially induced by an interaction/tidal disturbance. While spiral arms are generally thought to emerge in galaxies only at redshifts of $z\lesssim 2$ \citep{Elmegreen2014}, a handful of spiral galaxies have been claimed at higher redshift \citep[a three-armed spiral at $z=2.18$, and a one-armed spiral at $z=2.54$;][]{Law2012, Yuan2017}. Of the spirals, grand design ($m=2$) spirals, such as our observations suggest, are thought to extend the furthest back in time, likely due to the ability of interactions to drive such spirals. Specifically, tidal interactions from prograde encounters are very effective at inducing the formation of spiral arms, particularly of the $m$=2 variety \citep{Dobbs2014}. The perturber should ideally be 1/10 of the mass of the main galaxy to produce a clear grand design spiral pattern \citep{Oh2008}. 
Their apparent rarity at high redshift is likely due not only to the fact that specific circumstances must be achieved to incite the spiral pattern in the first place (the galaxy must be massive enough to have stabilized the formation of an extended disk, and the disk must then be perturbed by a sufficiently massive companion with the correct orientation), but also to the fact that such interaction-driven spirals are generally short-lived \citep[though this depends on the exact configuration and orbital parameters;][]{Law2012}. In that sense, and if this morphology is triggered by an interaction, it is perhaps not surprising that some of the ALESS SMGs show potential spiral structure, as they were selected through their bright submillimeter emission to be some of the most highly star-forming galaxies in the Universe, ensuring that they are both massive and viewed close in time after the presumed interaction.

\begin{deluxetable}{ l c c c }[]
\tabletypesize{\small}
%\tablewidth{10cm}
\tablecaption{Sources with potential bar$+$ring morphologies}
\tablehead{
\colhead{Source ID} & \colhead{`Bar' radius$^a$} & \colhead{`Ring' radius$^b$} & \colhead{Ratio (OLR/CR$^c$)} \\
 & [kpc] & [kpc] & -- }
 \startdata
 ALESS 15.1 & 0.7$\pm$0.1 & 1.9$\pm$0.1 & 2.6$\pm$0.2\\
 ALESS 17.1 & 1.0$\pm$0.1 & 1.9$\pm$0.1 & 1.5$\pm$0.2\\
 ALESS 76.1 & 0.7$\pm$0.1 & 1.3$\pm$0.1 & 1.9$\pm$0.2
 \enddata
 \tablecomments{$^a$Defined as the HWHM of the central (`bar') component along the major axis from a two-dimensional Gaussian fit.\\
 $^b$Defined as the average distance from the source center to the `clumps' bracketing the central emission along the major axis.\\
 $^c$The ratio of the `ring' and `bar' sizes, taken here to indicate the ratio of the outer lindblad resonance (OLR) and the corotation radius (CR). See Section~\ref{morphology} for further details.
 }
 \label{tab:barring}
 \end{deluxetable}

Alternately, the spiral structure visible in ALESS 112.1 -- which is also the source with the lowest S\'ersic index -- may instead be due to a late-stage major merger viewed at a serendipitous angle. 
The maximum starburst (and heaviest dust obscuration) coincides with final coalescence in retrograde-retrograde mergers, which also show appreciably larger internal dust extinction than prograde-prograde configurations \citep{Bekki2000}. The fact that the strongly star-forming component is on average more compact than both the gas and existing stellar component in SMGs \citep{Simpson2015b, Chen2017, CalistroRivera2018} would also be consistent with this picture \citep[e.g.,][]{Bekki2000}.  

In at least three of the sources (ALESS 15.1, 17.1, 76.1), we detect clump-like structures along the major axis of the ALMA emission, and bracketing elongated nuclear emission. This could suggest that we are observing bars in the cores of these galaxies, where the aligned clump-like structures are either star-forming gas complexes such as those frequently seen in local barred galaxies (e.g., NGC 1672) and which may be formed through orbit crowding in a bar-spiral transition zone \citep[e.g.,][]{Kenney1991, Kenney1992}, or they are due to a star-forming ring that is visible as two clumps due to the long path length where the line-of-sight is perpendicular. As the three sources with the strongest evidence for this morphology are also the most highly inclined sources based on the {\sc galfit} modeling, this could be evidence for the latter (a bar and ring morphology). 

Bars observed in the local Universe in the near-infrared are usually well fit by S\'ersic models with $n \sim 0.5-1$ \citep{Weinzirl2009}. It is difficult to separate the potential bars from the disks in a multi-component S\'ersic fit with the data presented here, so we can only say that these values are typically lower than what we measure for our galaxies on the whole.
If we define the radius of the bar as the HWHM of the central component along the major axis from a two-dimensional Gaussian fit, then we find bar radii of 0.7-1 kpc (Table~\ref{tab:barring}). 
 In \citet{Hoyle2011}, a Galaxy Zoo project which measured bar lengths in $\sim$3000 local SDSS disk galaxies, they report bar radii of $\sim$1-10 kpc, which would put these high-redshift ``bars'' on the extreme short end. (The rings are correspondingly small for what you would then expect based on the bar resonances -- see below). However, the same study finds that the shortest bars are found in the bluest (and thus presumably most star-forming) disk galaxies. This may be due to the gas content of the galaxies, as recent high-resolution hydrodynamic simulations of barred galaxies by \citet{Seo2019} find that higher gas fractions result in shorter bars, particularly in dynamically `cold' systems (based on the ratio of the radial and vertical velocity dispersions). In addition, based on the typical gas depletion timescales estimated for SMGs \citep[$\sim$100-200 Myr;][and references therein]{Casey2014}, it is possible that we could be observing bars shortly after their formation, when they are expected to be shorter \citep[e.g.,][]{Seo2019}. The comparison is complicated by the fact that bar formation is a highly nonlinear process, especially in the presence of gas, and depends sensitively on the initial galaxy parameters \citep{Seo2019}. It is interesting to note that all of the bars formed in the \citet{Seo2019} hydrodynamic simulations eventually thicken and form bulges.

If our identification of these features is correct, and if we assume that the bar extends approximately to the corotation radius (CR) in these galaxies \citep{Tremaine1984, Sanders1980, Lindblad1996, Weiner2001, Buta1986, Athanassoula1992, Perez2012},  
then the extent of the bar can also give us the corotation radius. In such a scenario, the rings form due to gas accumulation at the bar resonances, 
and the diameter of the rings gives the outer Lindblad resonance (OLR). Taking these three galaxies (ALESS 15.1, 17.1, and 76.1), we define the radius of the bar as above and the radius of the `ring' as the average distance to the `clumps' from the source center, resulting in a median ratio of the two sizes (interpreted here as OLR/CR) of 1.9$\pm$0.3. This ratio agrees with the OLR/CR ratio found for the local galaxy population\footnote{Although we note that the morphological characteristics of the bar region of galaxies are strongly influenced by properties of the ISM which may differ at high-redshift, such as gas fraction \citep[][]{Athanassoula2013}.} \citep[e.g.,][]{Kormendy1979b, Buta1995, Laurikainen2013, HerreraEndoqui2015}, supporting this interpretation of these features. Notably, this also agrees with the theoretical prediction from density wave theory for the assumption of a flat rotation curve in the inner disk. For galaxies with rising rotation curves, the OLR of the (stellar) 
bar would be spaced further from its edge \citep{MunozMateos2013}.\footnote{Note also that, contrary to the long-standing belief, recent hydrodynamical simulations show that the presence of a stellar bar does not imply that baryons dominate gravitationally in that region \citep{Marasco2018}.}

Galaxies with bars are very common in the local Universe, with almost two-thirds of nearby galaxies classified as barred in infrared images that trace the stellar population \citep[e.g.,][]{deVaucouleurs1991, Knapen2000, Whyte2002, Laurikainen2004, MenendezDelmestre2007b, Buta2015, Laine2016}. The decline in the barred fraction of disk galaxies from $f_{\rm bar}$ $\sim$0.65 at $z=0$ to $f_{\rm bar}$ $<$0.2 at $z = 0.84$ \citep{Sheth2008} is almost exclusively in the lower-mass (M$_{*}$ $=$ 10$^{10-11}$ M$_{\odot}$), later-type, and bluer galaxies, potentially due to their dynamically hotter disks \citep{Sheth2012}. In more massive, dynamically colder disks, studies have shown that bars can form out to high redshift \citep[$z\sim1-2$;][]{Jogee2004, Simmons2014}.
While bars can occur without an interaction, the tidal forces induced by interactions have long been suspected to play a role in bar formation \citep[e.g.,][]{Elmegreen1990, Athanassoula2002}, particularly for massive galaxies \citep{Mendez-Abreu2012},
and bars and rings are frequently found together in local interacting systems. Thus, if this interpretation is correct, this could be another indication of interaction-induced substructure in these SMGs. Indeed, the presence of a bar itself would indicate an unstable disk; i.e., a Toomre stability parameter 
\begin{equation}
Q = \frac{\sigma_{\rm r}\kappa}{\pi\rm{G}\Sigma_{\rm disk}} < 1
\end{equation}
where $\sigma_{\rm r}$ is the one-dimensional velocity dispersion, $\kappa$ is the epicyclic frequency, and $\Sigma_{\rm disk}$ is the surface density of the disk. Here we assume that the gas disk dominates over the stellar component. Taking the epicyclic frequency appropriate for a flat rotation curve \citep[$\kappa = \sqrt{2}V_{\rm max}/R$ with an assumed $V_{\rm max}$ = 300 km s$^{-1}$ as typical for SMGs;][]{Bothwell2013}, taking the radius as the HWHM of the ALMA 870\,$\mu$m continuum emission along the major axis, and again estimating the gas surface density from the global $\Sigma_{\rm SFR}$, we derive upper limits for the one-dimensional velocity dispersion of the potentially barred sources of $\sigma_{\rm r} \lesssim$ 70--160 km s$^{-1}$ (Table~\ref{tab:veldisp}). For these three sources, these Toomre-based upper limits are consistent, albeit marginally, with the lower limits derived from equating the sizes of the largest `clumps' observed to the Jeans length.

If we are indeed observing bar$+$ring and spiral arm morphologies in some of the sources, we note that the velocity fields would have crossing orbits which would allow efficient loss of angular momentum and collisionally induced star formation. These non-axisymmetric structures (particularly bars) force the gas streams to cross and shock, increasing star formation efficiency and allowing for net angular momentum loss \citep[e.g.,][]{Hopkins2011}. The observations presented here may therefore be uncovering the detailed physical mechanisms which result in the very high SFRs measured for SMGs. Ultimately, high-resolution kinematic information is necessary to test the various physical interpretations and confirm the values of the relevant parameters discussed above.

%%%%%%%%%%%%%%%%%%%%%%%%%%
\section{Summary}
\label{summary}	

We have presented high--fidelity 0.07$''$ imaging of the 870\,$\mu$m continuum emission in six luminous
galaxies ($z=1.5-4.9$) from the ALESS SMG survey, allowing us to map the rest-frame FIR emission on $\sim$500\,pc scales. 
Our findings are the following:

\begin{itemize}

\item We report evidence for robust sub-kpc structure on underlying exponential disks. These structures have deconvolved sizes of $\lesssim$0.5--1\,kpc. They collectively make up $\sim$2--20\% of the total continuum emission from a given galaxy, indicating they are not the dominant sites of star formation (assuming a constant dust temperature).

\item We observe no correlation between these structures and those seen in lower-resolution {\it HST} imaging, which is extended on larger scales. This comparison suggests that we may be probing the heavily dust-obscured cores of more extended systems. 

\item The large-scale morphologies of the structures show (1) clear curvature in the inner $\sim$5--10\,kpc for two galaxies (ALESS 112.1 and 15.1), and (2) pairs of clump-like structures bracketing elongated nuclear emission in the three sources that appear to be the most edge-on (ALESS 15.1, 17.1, and 76.1). These observations are suggestive of (1) spiral arms, and (2) bars and star-forming rings in inclined disks. The ratio of the `ring' and `bar' radii (1.9$\pm$0.3) is consistent with local galaxies, lending support to this interpretation. The presence of such features may be an indication of tidal disturbances in these systems. 

\item We use our high-resolution 870\,$\mu$m imaging to create maps of the star formation rate density ($\Sigma_{\mathrm{SFR}}$) on $\sim$500\,pc scales within the sources, finding peak values that range from $\sim$40--600 M$_{\odot}$ yr$^{-1}$ kpc$^{-2}$ between sources. We trace this large range in peak $\Sigma_{\mathrm{SFR}}$ back to different galaxy-integrated physical conditions (dust luminosities and temperatures) in the galaxies. 

\item Compared to a sample of local U/LIRGs, the SMGs appear to be able to sustain high (galaxy-averaged) rates of star formation over much larger physical scales. However, even on 500\,pc scales, they do not exceed the Eddington limit set by radiation pressure on dust. 
The peak $\Sigma_{\mathrm{SFR}}$ values measured are consistent with those seen in U/LIRGs with similar (total) sizes. As local U/LIRGs can achieve even higher $\Sigma_{\rm SFR}$ values on smaller physical scales than observable in the SMGs, this may indicate that higher resolution ($<$500 pc FWHM) observations would be necessary to observe super-Eddington star formation in typical SMGs.

\end{itemize}

Further observations are required to verify the results presented here. In particular, resolved multi-frequency continuum mapping with ALMA is necessary to constrain the variation in dust temperature within the sources (which would affect the derived $\Sigma_{\mathrm{SFR}}$ maps), and a larger sample size is important for moving beyond the handful of submillimeter-brightest sources studied here. The striking comparison with the \textit{HST} imaging highlights the need for high-resolution, near-IR imaging of such dusty targets, such as will become possible with \textit{JWST}. 
Finally, high-resolution kinematics are key for confirming the existence of non-axisymmetric structures within inclined disks.
If confirmed by kinematics, the presence of bars would imply that the galaxies have flat rotation curves and Toomre-unstable disks ($Q<1$). The implied one-dimensional velocity dispersions ($\sigma_{\rm r} \lesssim$ 70--160 km s$^{-1}$) would be marginally consistent with the lower limits suggested from equating the sizes of the largest clump-like structures observed to the Jeans length.
Finally, such non-axisymmetric structures would provide a mechanism for net angular momentum loss and efficient star formation, helping to explain the very high SFRs measured in SMGs.

%%%%%%%%%%%%%%%%%%%%%%%%%%%%%%%%%%%%%%%%%%%%
\acknowledgements
The authors wish to thank the anonymous referee for helpful comments which improved this paper.
We also thank Sharon Meidt, Arjen van der Wel, Aaron Evans, Francoise Combes, Rob Ivison, and Karin Sandstrom 
for useful discussions and advice. 
JH and MR acknowledge support of the VIDI research program with project number 639.042.611, which is (partly) financed by the Netherlands Organization for Scientific Research (NWO). 
IRS acknowledges support from STFC (ST/P000541/1) and the ERC Advanced Grant {\sc dustygal} (321334). 
EdC gratefully acknowledges the Australian Research Council for funding support as the recipient of a Future Fellowship (FT150100079).
HD acknowledges financial support from the Spanish Ministry of Economy and Competitiveness (MINECO) under the 2014 Ram\'on y Cajal program MINECO RYC-2014-15686.
JLW acknowledges support from an STFC Ernest Rutherford Fellowship (ST/P004784/1 and ST/P004784/2).
This paper makes use of the following ALMA data: ADS/JAO.ALMA\#2016.1.00048.S, and ADS/JAO.ALMA\#2011.1.00294.S. 
ALMA is a partnership of ESO (representing its member states), NSF (USA) and NINS (Japan), together with NRC (Canada) and NSC and ASIAA (Taiwan) and KASI (Republic of Korea), in cooperation with the Republic of Chile. The Joint ALMA Observatory is operated by ESO, AUI/NRAO and NAOJ.
The National Radio Astronomy Observatory is a facility of the National Science Foundation operated under cooperative agreement by Associated Universities, Inc.

%%%%%%%%%%%%%%%%%%%%%%%%%%%%%%%%%%%%%%%%%%%%%
% BIBLIOGRAPHY

\bibliographystyle{apj}		% Make sure apj.bst is in your current directory
%\bibliography{../MyRefs}
\bibliography{ALESScyc4_hires.bib}

%%%%%%%%%%%%%%%%%%%%%%%%%%%%%%%%%%%%%%%%%%%%%%%
% LONG FIGURES

%%%%%%%%%%%%%%%%%%%%%%%%%%%%%%%%%%%%%%%%%
% LONG TABLES

\end{document}